\documentclass[11pt, preprint]{aastex}

\usepackage{geometry}                
\usepackage{graphicx}
\usepackage{amssymb}
\usepackage{epstopdf}
\usepackage{longtable, lscape}

\newcommand{\Msun}{\hbox{M$\sb{\odot}$}}

\begin{document}

\title{Rocky Extrasolar Planetary Compositions Derived from Externally-Polluted White Dwarfs}


\author{B. Klein\altaffilmark{1}, M. Jura\altaffilmark{1}, D. Koester\altaffilmark{2}, B. Zuckerman\altaffilmark{1}}

\altaffiltext{1}{Department of Physics and Astronomy, University of California, Los Angeles, CA 90095-1562, USA; kleinb, jura, ben@astro.ucla.edu}
\altaffiltext{2}{Institut fur Theoretische Physik und Astrophysik, University of Kiel, 24098 Kiel, Germany; koester@astrophysik.uni-kiel.de}

\begin{abstract}

We report Keck High Resolution Echelle Spectrometer data and model atmosphere analysis of two helium-dominated white dwarfs, PG1225-079 and HS2253+8023, whose heavy pollutions most likely derive from the accretion of terrestrial-type planet(esimal)s.  For each system, the minimum accreted mass is $\sim$10$^{22}$ g, that of a large asteroid.  In PG1225-079, Mg, Cr, Mn, Fe and Ni have abundance ratios similar to bulk Earth values, while we measure four refractory elements, Ca, Sc, Ti and V, all at a factor of $\sim$2--3 higher abundance than in the bulk Earth.   For HS2253+8023 the swallowed material was compositionally similar to bulk Earth in being more than 85\% by mass in the major element species, O, Mg, Si, and Fe, and with abundances in the distinctive proportions of mineral oxides -- compelling evidence for an origin in a rocky parent body.  Including previous studies we now know of four heavily polluted white dwarfs where the measured oxygen and hydrogen are consistent with the view that the parents bodies formed with little ice, interior to any snow-line in their nebular environments.  The growing handful of polluted white dwarf systems with comprehensive abundance measurements form a baseline for characterizing rocky exoplanet compositions that can be compared with bulk Earth.  

\end{abstract}


\maketitle

\section{INTRODUCTION\label{sec:intro}}

Accretion by white dwarfs of tidally disrupted bodies from their planetary systems \citep{jura2003} has become the most viable explanation for the presence of high-Z (atomic number Z$>$2) elements in the vast majority of polluted DAZ, DBZ, and DZ white dwarf atmospheres with T$_{eff}$ $<$ 20,000 K \citep[][and references therein]{jura2008, zuckerman2010, farihi2010mnras}.  
The basic idea is that smaller planetary bodies such as asteroids or dwarf planets, can acquire highly eccentric orbits by perturbations from larger planetary bodies, especially considering the dynamical rearrangement of the planetary system during stellar evolution and mass loss \citep{debsig2002}.   \citet{bonsor2011arXiv} demonstrate that the post-stellar mass loss evolution of a planetesimal belt can supply the required mass influx to explain observed polluted white dwarf (WD) accretion rates.    Objects that journey within the Roche radius of the WD will experience tidal shredding, forming a disk of dust and/or gas \citep{jura2003,jura2008}, and subsequently accrete into the star's, otherwise pure, hydrogen and/or helium atmosphere.   This natural mechanism offers the potential to measure the distilled elemental constituents of a planetary parent body or bodies -- a unique and powerful tool, indeed.  

High-resolution and high-sensitivity spectroscopy, along with state-of-the-art model atmospheres \citep{koester2009,koester2010, tremberg2009, dufour2010, vennes2010, vennes2011}, has opened the door to detailed studies of WD atmospheres and the high-Z material that pollutes them.  To date, a handful of WDs with very rich spectra have been studied at high-resolution and high-sensitivity, beginning with the Keck/HIRES follow-up \citep{zuckerman2007} of GD 362, whose extreme high-Z-rich atmosphere was pointed out by \citet{gianninas2004} and \citet{kawkavennes2006gd362}.   The detection of oxygen (O) along with the other major elements (Mg, Si, and Fe) in GD40's polluting parent body \citep{klein2010} established chemical evidence  -- the stoichiometric balance of mineral oxides -- for rocky extrasolar planetesimals.  Ongoing discoveries of polluted WDs displaying all major terrestrial elements \citep{vennes2010, dufour2010, zuckerman2010, farihi2011gd61} signify that externally polluted white dwarfs have entered a stage of being practical measuring devices of the elemental compositions of rocky exoplanets.  \citet{bond2010} predict a wide diversity of terrestrial exoplanets including Earth-like, refractory-rich, and carbon-rich objects.  Polluted white dwarfs provide real data at the level of detail needed to make discerning measurements \citep[e.g.,][]{zuckerman2011}.

 In this paper we report the detailed analysis of HIRES spectra of HS2253+8023 (V$_{mag}$=16.1, DBAZ, T$_{eff} \sim$ 14,400 K; hereafter HS2253) and PG1225-079 ($\equiv$K 789-37, V$_{mag}$=14.8, DZAB, T$_{eff} \sim$ 10,800 K; hereafter PG1225).  Both stars were previously known to display features of H, Mg, Si, Ca and Fe mainly from low ($\sim$ 6\AA) resolution IUE and optical data \citep{koester1990, friedrich1999, wolff2002}.  Calcium in PG1225 was also (re)observed and analyzed from the ESO SN Ia progenitor (SPY) survey \citep{koester2005}.  The HIRES observations provide a new and important detection of oxygen in HS2253, completing the set of major terrestrial elements -- O, Mg, Si, and Fe -- as well as H, Ca, Ti, Cr, Mn and possibly, Ni.  For PG1225 the HIRES data display six newly identified elements:  Sc, Ti, V, Cr, Mn, and Ni.  By analyzing all the detected elements together in a given model, with multiple lines for each element, and accounting for abundance sensitivity to errors of the model parameters, we calculate pollutant-to-pollutant abundance ratios with an estimated uncertainty of $\sim$ 40\% or less, a significant improvement over previous analyses in which derived abundances relative to helium are typically uncertain by 0.2-0.7 dex.  Infrared studies show that HS2253 does not have an infrared excess \citep{farihi2009}, and PG1225 has just a subtle infrared excess appearing only at 7.9$\micron$ \citep{farihi2010, kilic2008dust}.  Notwithstanding the lack of detected circumstellar dust, we demonstrate that these systems have accreted an appreciable amount of material with terrestrial planet(esimal) composition.

We also report the detection of Ca {\small II} H\&K emission cores in PG1225, the first time seen in a white dwarf.  Other peculiar line profiles observed with HIRES are narrow absorption cores in helium lines of PG1225, GD362 and GD16, which are displayed in this paper.

There are a number of benefits of working with helium-dominated WDs.  One is that, compared to hydrogen-dominated WD atmospheres, the lower opacity helium environments more readily display tiny concentrations of numerous contaminant elements, e.g.\  the trace elements vanadium and scandium are detected in PG1225 at 10 - 11 orders of magnitude lower abundance than the helium background they are diluted in.  Secondly, detailed model atmosphere analysis of helium-atmosphere WDs indicates the hydrogen in these stars must be accreted \citep{voss2007,dufour2007}; thus when helium is the background constituent, one obtains a measurement of the polluting hydrogen abundance, which is important for understanding the potential accretion of water ice \citep{jura2009apj,juraxu2010}.   A simplifying aspect of the helium-atmosphere WDs is that in our T$_{eff}$ range of interest ($<$ 20,000 K), all have outer convection zones where the elements are homogeneously mixed.   Finally, another benefit of working with helium-rich WDs is that with relatively large convective zones (relatively long settling times), the pollutants are not disappearing so fast from our view.  Thus, a higher fraction of polluted WDs can be found \citep[e.g., ][]{zuckerman2010}, and meaningful lower limits on accreted masses can be obtained.  Recently, \citet{dufour2010} and \citet{koester2011} have established that absolute minimum accreted masses for a number of polluted WDs are $\gtrsim$10$^{23}$ g, close to that of Ceres, the largest asteroid in our own Solar system.


Gravitational settling times are always less than the age of the WD, but in helium-atmosphere stars of T$_{eff}$ $<$ 20,000K, the settling (diffusion) time scales for most elements are 10$^3$-10$^7$ years \citep{koester2009}.  Thus a measurement of atmospheric abundances in a helium-dominated WD is a snapshot of one stage of the evolving interplay of accretion and diffusion in that system.  As was found for  GD40 \citep{klein2010}, the patterns of pollutant abundances themselves, can narrow the range of possibilities for the accretion--diffusion state of a WD, and we utilize some such constraints in the present work.

The article is laid out as follows: observations, data reduction, and measurements of the spectra are described in Section \ref{sec:obs_meas}.  In Section \ref{sec:models} we discuss effective stellar temperature, surface gravity, and the synthetic model atmospheres from which we derive the polluting element abundances presented in Section \ref{sec:abundances}.   We describe the  effects of accretion and diffusion in Section \ref{sec:accretiondiffusion}, incorporate that into our results -- including results of previous studies -- in Section \ref{sec:resultsdiscussion}, and give our conclusions in Section \ref{sec:conclusions}.

\section{OBSERVATIONS \& MEASUREMENTS\label{sec:obs_meas}}

The data were acquired with the High Resolution Echelle Spectrometer \citep[HIRES,][]{vogt1994} on the Keck I telescope of Mauna Kea Observatory.  Table \ref{tab:obs} is the observation log. Typically, a 1\farcs148 slit width was used, providing a spectral resolution of $R = \lambda/\Delta\lambda \sim 40,000$.  To compensate for inferior seeing on 2008 Nov 15, a 1\farcs722 slit was used, resulting in a resolution of $R \sim 28,000$ for those spectra.  Before combining exposures of the same star with differing setups, the higher resolution spectra were degraded by Gaussian convolution to approximate the lower resolution spectra. 

\begin{figure}[htbp]
\begin{center}
  \includegraphics[width=140mm]{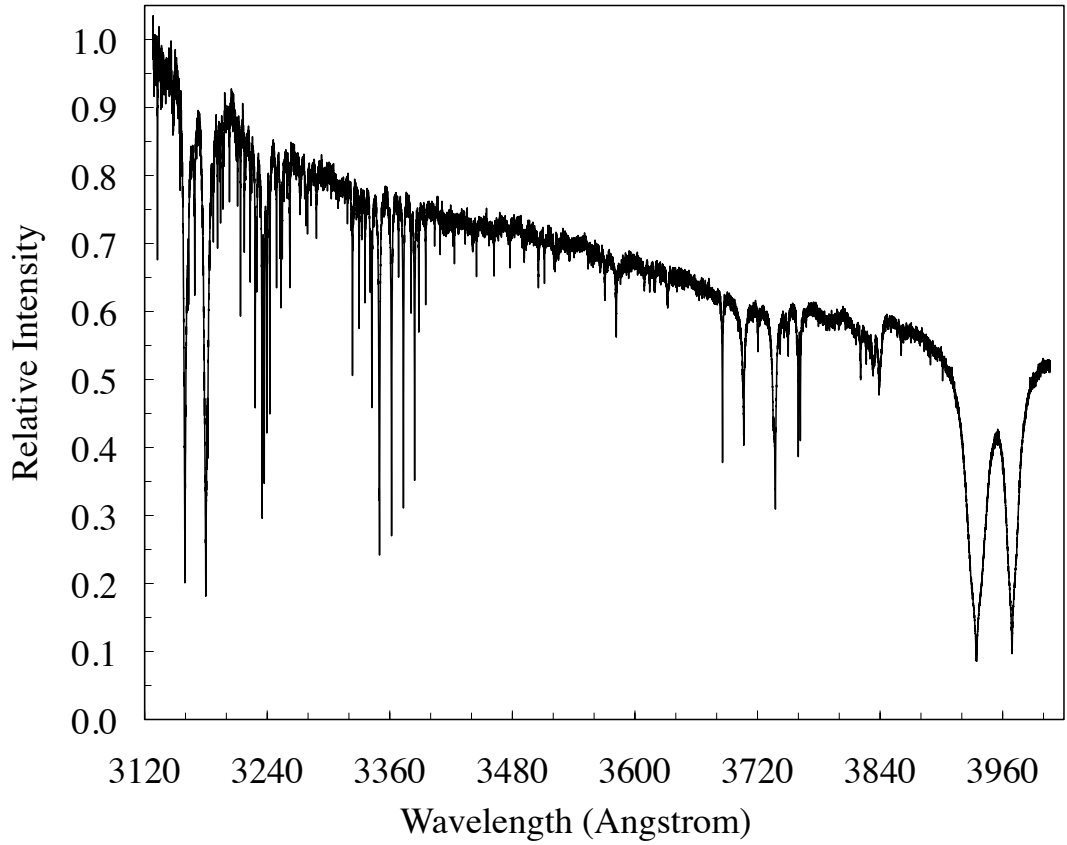}
\caption{Combined, echelle-order-merged, shortest wavelength portion (CCD1 with blue collimator) of HIRES spectrum of PG1225-079.  The number density of absorption lines increases dramatically toward UV wavelengths.  The absolute flux level and the overall continuum slope are uncalibrated, which is not a problem for the line measurements which are always made relative to their local continuum.}
\label{fig:pg1225_blu}
\end{center}
\end{figure}

\begin{figure}[htbp]
\begin{center}
  \includegraphics[width=140mm]{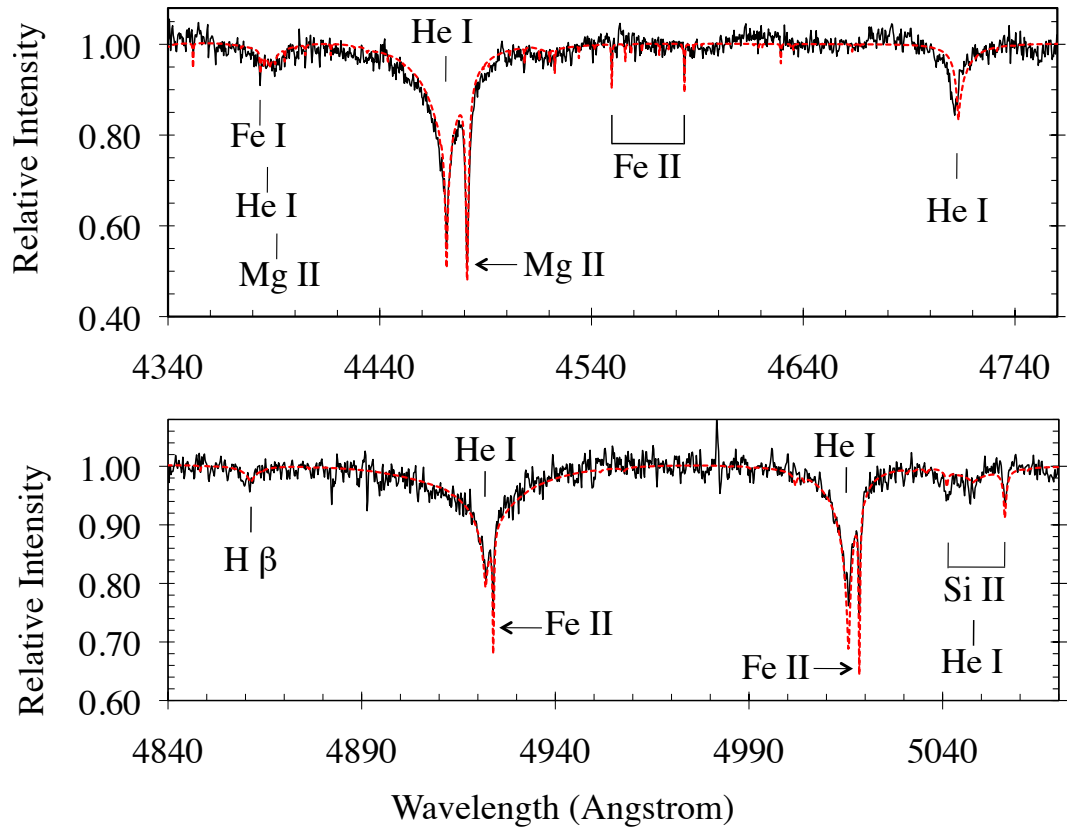}
\caption{Combined, echelle-order-merged HIRES spectrum of HS 2253+8023 (black), highly smoothed for clarity in displaying the broad helium features over the large wavelength range plotted.   A T$_{eff}$=14,400 K, log $g$ = 8.4 model, with element abundances from Table \ref{tab:hs2253_abund}, is plotted in red. }
\label{fig:hs2253_He}
\end{center}
\end{figure}

Data reduction and absorption line measurements were carried out using PyRAF in a similar manner as described for GD40 in \citet{klein2010}.  For the present work we took some extra steps, since in addition to a fully continuum-normalized version of each spectrum, we also wanted a version that preserved broad-line ($>$ 1-2 \AA) features.  For this purpose we used observations of calibration white dwarfs (BD+28 4211, G191-B2B, or Feige 34) in the flux-calibration procedure to align the echelle orders (however, absolute flux calibration is not done).  As is known for HIRES data \citep{suzuki2003}, echelle artifacts irregularly appear as bends or waves in the continuum.  We visually inspected and compared regions of similar wavelength coverage, and selectively combined only those portions of the spectra that appeared not to suffer significant distortions.  Mostly this amounted to omitting some echelle-order edges.  Figures \ref{fig:pg1225_blu} and \ref{fig:hs2253_He}  show the high-quality output, and success with broad lines, from these procedures.   Residual artifacts are low-level and fairly broad, which for example can be seen as waves in the continuum in the region $\lambda$4550-4680\AA\ of Figure \ref{fig:hs2253_He}.

For HS2253 the continuum signal-to-noise (SNR) of the data ranges from 25--40 for wavelengths in 3400--8000\AA, and is 10--15 at wavelengths shorter and longer than that.  
The spectrum of PG1225 is very high quality with the SNR of the unsmoothed continuum ranging from 80--110 for wavelengths in 3200--6000\AA\, and 40--70 at wavelengths shorter and longer than that.

\begin{figure}[htbp]
\begin{center}
  \includegraphics[width=140mm]{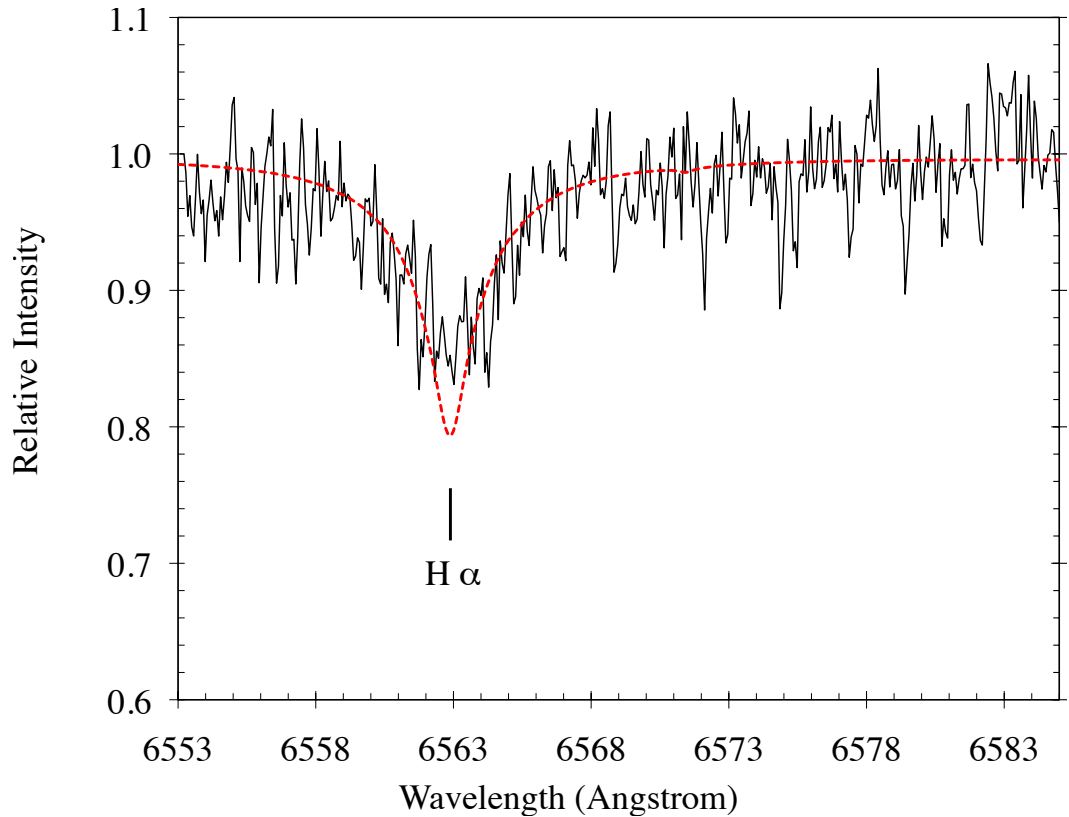}
\caption{Portion of the HIRES data (black) of HS2253+8023, smoothed by a 5-point boxcar average and displaying H$\alpha$.   Wavelengths are in air and a heliocentric rest frame.   A T$_{eff}$=14,400 K, log $g$ = 8.4 model, convolved with a Gaussian to approximate the instrumental resolution and with the hydrogen abundance from Table \ref{tab:hs2253_abund}, is plotted in red. The echelle order end occurs in the data near 6553\AA, thus the continuum blue-ward of the H$\alpha$ feature is minimally sampled, but is adequate for an EW measurement, which is matched with the model. 
}
\label{fig:hs2253_Halpha}
\end{center}
\end{figure}

\begin{figure}[htbp]
\begin{center}
  \includegraphics[width=140mm]{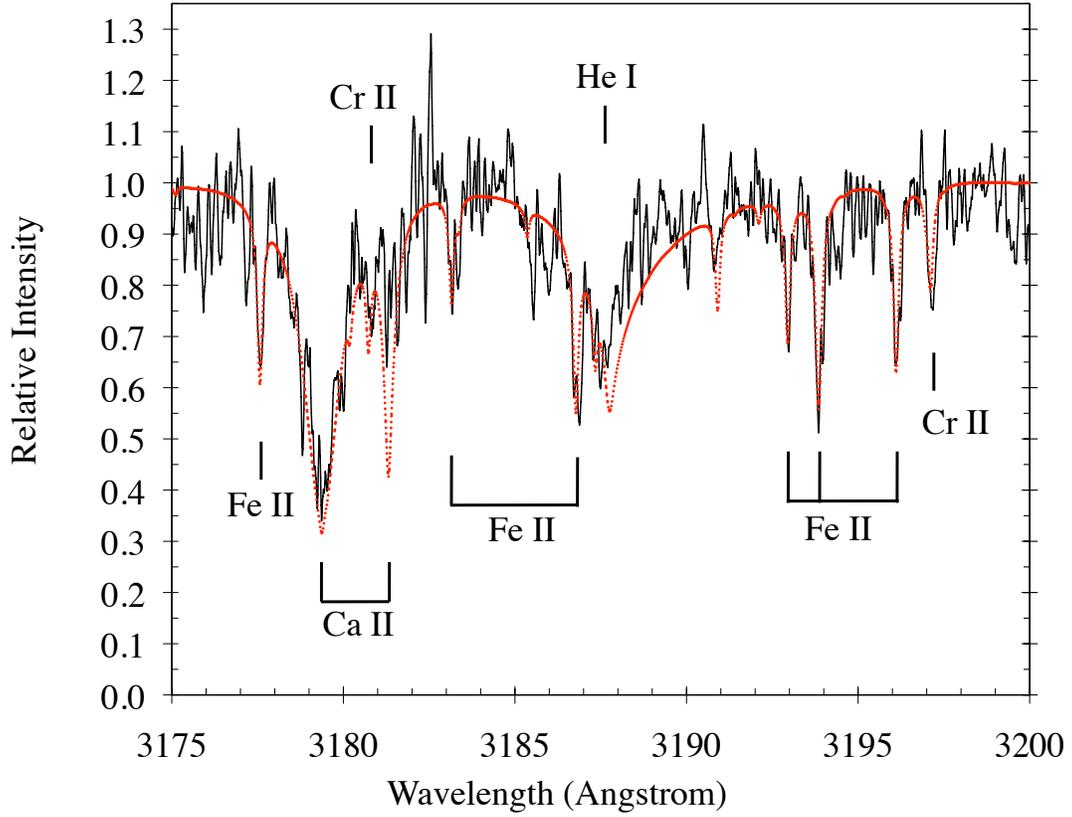}
\caption{Portion of the HIRES data (black) of HS2253+8023, smoothed with a 5 point boxcar average and displaying lines of He {\small I}, Ca {\small II}, Cr {\small II}, and Fe {\small II}.  Wavelengths are in air and a heliocentric rest frame.  A T$_{eff}$=14,400 K, log $g$ = 8.4 model, convolved with a Gaussian to approximate the instrumental resolution and with element abundances from Table \ref{tab:hs2253_abund}, is plotted in red. }
\label{fig:hs2253_3185}
\end{center}
\end{figure}

\begin{figure}[htbp]
\begin{center}
  \includegraphics[width=140mm]{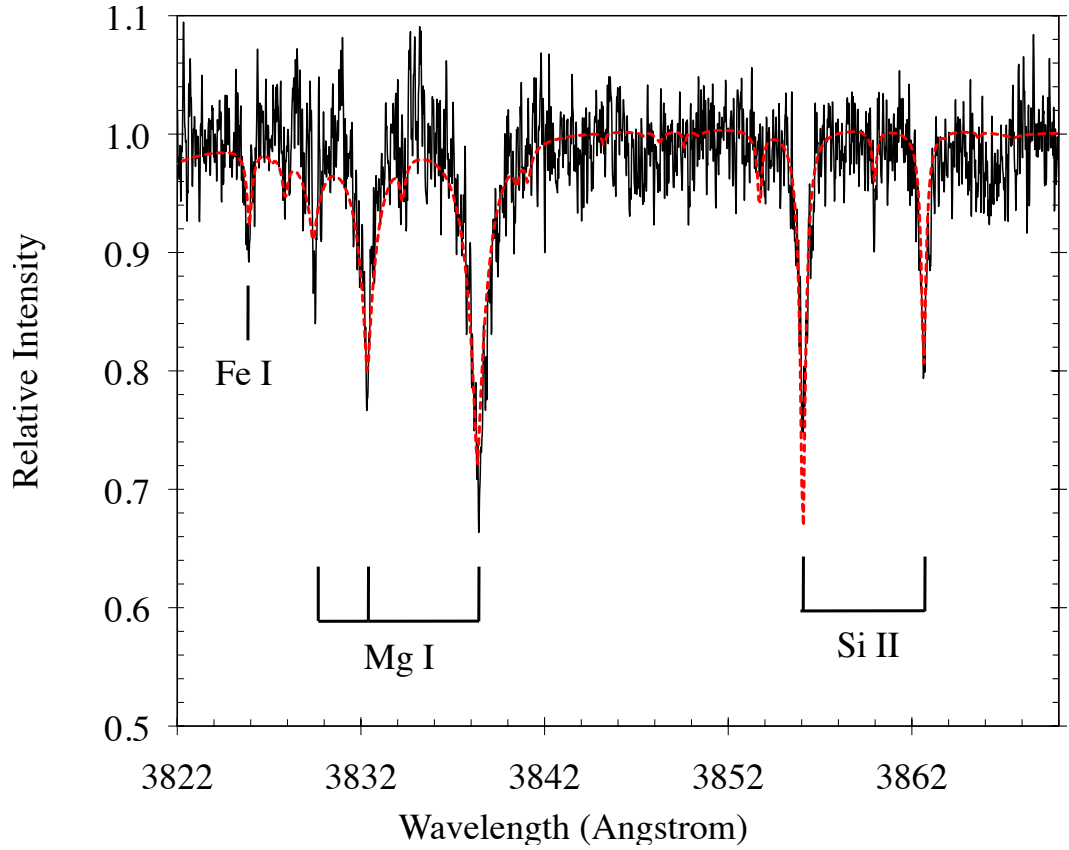}
\caption{HS2253+8023.  Similar to Figure \ref{fig:hs2253_3185}, but for Mg {\small I}, Si {\small II}, and Fe {\small I}.  This portion of the data was smoothed with a 3 point boxcar average.}
\label{fig:hs2253_MgSi}
\end{center}
\end{figure}

\begin{figure}[htbp]
\begin{center}
  \includegraphics[width=140mm]{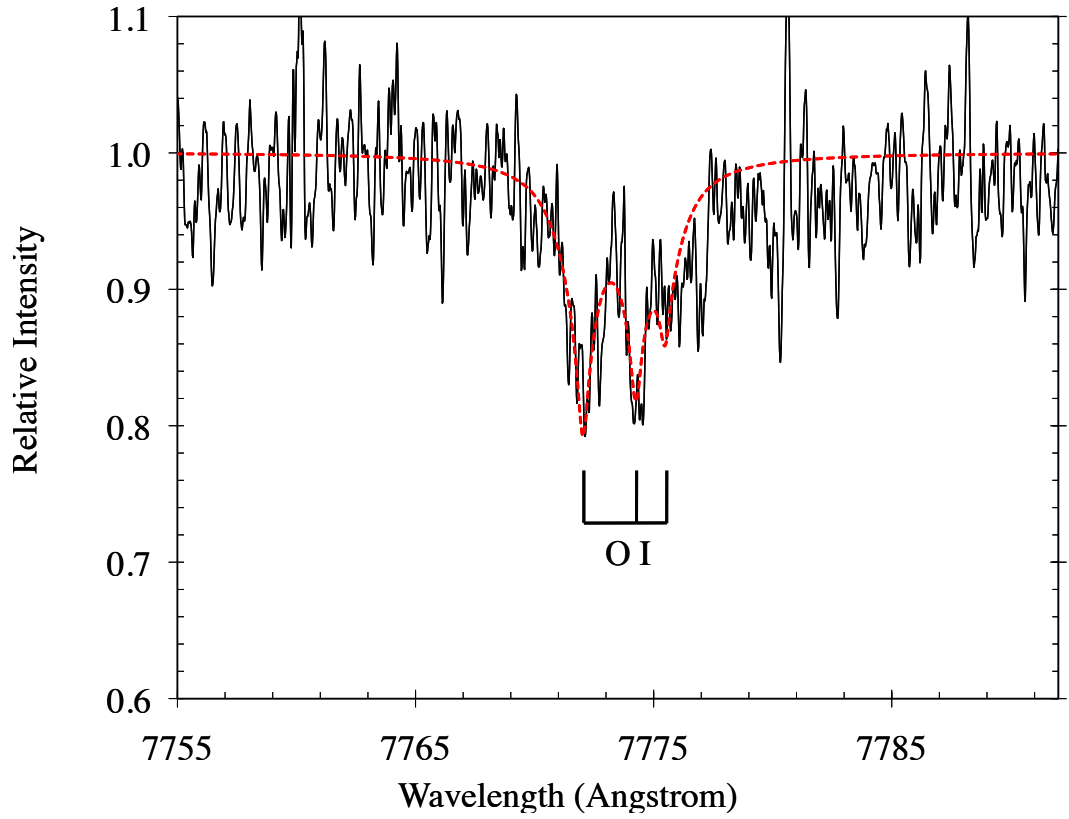}
\caption{HS2253+8023. Similar to Figure \ref{fig:hs2253_3185}, but for O {\small I}.  This portion of the data was smoothed with a 5 point boxcar average.}
\label{fig:hs2253_OI}
\end{center}
\end{figure}

\begin{figure}[htbp]
\begin{center}
 \includegraphics[width=140mm]{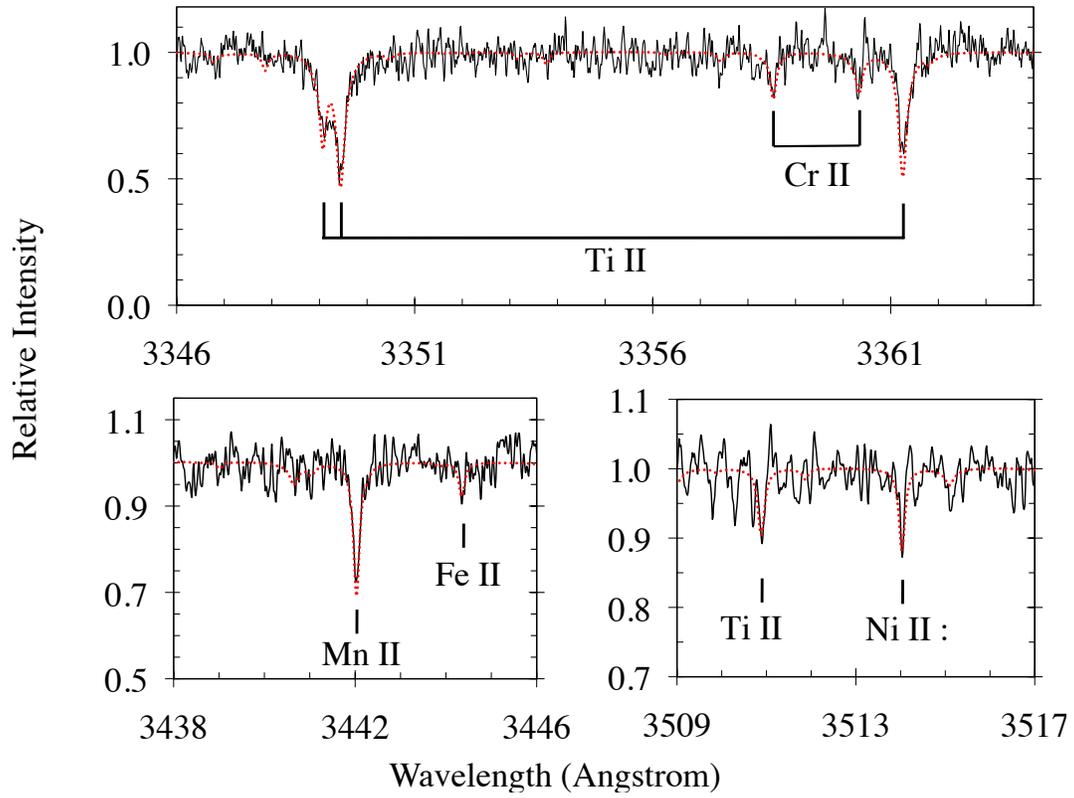}
\caption{HS2253+8023. Similar to Figure \ref{fig:hs2253_3185}, but displaying lines of Ti {\small II}, Cr {\small II}, Mn {\small II}, and a marginal detection of Ni {\small II}.}
\label{fig:hs2253_TiMnNi}
\end{center}
\end{figure}

The line identifications and equivalent widths are listed in Table \ref{tab:linelist}, and were carried out in a similar manner as described for GD40 in \citet{klein2010}.   For PG1225 the many broad lines and high number density of lines was not amenable to our usual method of measurement error estimation.  Instead, based on our experience with HIRES data and checks over multiple exposures, for PG1225 we set the EW measurement errors to 20\% for most lines, and 30\% or more for a few lines where the continuum or interval surrounding the line is especially uncertain.  We have also included here the details of the line measurements of G241-6 (SDSS g =15.5 mag, DBZ, T$_{eff}$=15,300 K, log $g$=8.0), whose high-Z rich nature and pollutant abundances were reported by \citet{zuckerman2010}.   Especially in the case of the very rich spectrum of PG1225, there are a number of blended lines, which are indicated with a `b' superscript in Table \ref{tab:linelist}.   In a few regions, such as the Mg {\small I} $\lambda$3835 triplet, instead of taking the EW from fitting a Lorentzian or Voigt profile, we report the EW from a numerical sum of the flux under the continuum for a given interval surrounding the features. 
Owing to the excellent SNR of these spectra, the measured line strengths span a remarkably large range -- a factor of 3000 -- from 0.011 \AA\ (V {\small II}), to $\sim$1 \AA\ (Ti {\small II}) to greater than 30 \AA\ (Ca {\small II} H\&K blend).  All observed absorption lines, and those not observed, are consistent with expectations of line strengths for our derived abundances and SNRs of the spectra.  Displays of various portions of the HIRES spectra are given in Figures \ref{fig:pg1225_blu}  -- \ref{fig:g241-6_FeMgSi}.

 For each transition, the shift of line center from the laboratory wavelength is measured, and converted to a heliocentric velocity.  We use the average line velocity for a systemic value, and Doppler shift the models to the measured frame of the WD, which are plotted in red on the figures of spectra.  The net systemic heliocentric velocities, including the stars' gravitational redshifts, are 4 $\pm$ 4 km s$^{-1}$ for HS2253 and 49 $\pm$ 3 km s$^{-1}$ for PG1225.  From the histogram of line velocity measurements shown in Figure \ref{fig:vel_hist}, for PG1225 we found a range of outliers (filled bars in the figure) that represent a detectable shift of the set of Fe {\small I} lines (see also Figure \ref{fig:pg1225_FeI_Sc}).  These are not of an interstellar origin, since as indicated in Table \ref{tab:linelist}, most transitions arise from excited atomic levels.  An analogous shift for Si {\small II} lines has been examined and modeled with Stark shifts by \citet{vennes2011}, thus it is conceivable that a similar process is responsible for the effect on Fe {\small I} in PG1225; further investigation remains for future work.  HS2253's velocity measurements do not indicate any distinctive sub-populations.

With such high-sensitivity observations of PG1225 a new feature has turned up (literally) that we have not seen before in WDs -- emission cores in the Ca {\small II} H\&K lines as shown in Figure \ref{fig:pg1225_CaHK}.  The emission features are seen in all 3 separate exposures covering the Ca H\&K region of PG1225, with no noticeable variations between exposures.   Emission cores have been seen in H$\alpha$ of hydrogen-dominated white dwarfs hotter than 25,000K, and are understood to arise from non-local-thermodynamic-equilibrium (NLTE) effects.  The one HIRES exposure covering H$\alpha$ in PG1225, taken about 2 weeks later, does not display clear H$\alpha$ core emission with a continuum SNR$\sim$50 near 6650\AA.  Still, NLTE effects could conceivably cause emission in the strong Ca H\&K lines since the cores are formed high up in the atmosphere, while the relatively weaker H$\alpha$ line of PG1225 is likely formed at a deeper location.  Alternatively, Ca H\&K emission cores, as well as H$\alpha$ core emission, are well known signatures of chromospheric activity in main sequence stars; further discussion of chromospheric activity in a WD is beyond the scope of this work \citep[see][]{musielak2005}.

One more spectroscopic peculiarity we point out is the presence of a narrow absorption core component at the line center of He {\small I} $\lambda$5876 in PG1225, as well as in GD362 and GD16 ($\equiv$ HS 0146+1847) shown in Figure \ref{fig:3He5876cores}.  These three WDs have helium-dominated atmospheres, with effective temperatures in the range 10,500 K -- 11,500 K;  they all have detected circumstellar dust \citep{becklin2005, kilic2005, farihi2009}, heavy atmospheric pollution \citep{koester2005gd16, zuckerman2007}, and possess relatively large amounts of hydrogen in their convective envelopes.  Two possibly analogous, but weakly detected, features in PG1225 are narrow absorption lines instead of broad ones centered on He {\small I} $\lambda$3889 and He {\small I} $\lambda$4472, with the latter shown in Figure \ref{fig:pg1225_He5876}.   
The He {\small I} $\lambda$5876 core feature was first discovered in HIRES data of GD362 by \citet{zuckerman2007} and noted again by \citet{tremblay2010};  we now see it in HIRES spectra of GD16 and PG1225 as well.  \citet{zuckerman2007} suggested that NLTE effects, similar to those seen in the cores of H$\alpha$ lines in hydrogen-dominated white dwarfs, could be present.   The exact origin of the He core feature is not known at this time and we do not discuss it further in this paper.

\section{MODEL ATMOSPHERES}
\label{sec:models}

The model input physics is described by \citet{koester2010} with all absorption line data taken from the Vienna Atomic Line Database (VALD). Key input parameters are the effective temperature, T$_{eff}$, surface gravity, $g$, and the atmospheric abundances of elements relative to the dominant constituent, in the current study, helium.  In the range of temperatures considered here, helium is mostly neutral, and the high-Z elements and hydrogen are significant contributors of electrons in the atmosphere.  Thus, we calculated and analyzed our models with hydrogen and high-Z elements in a self-consistent way, and included Earth-like abundances for undetected but expected elements (such as O in PG1225).  More than 2000 continuous absorption coefficients were incorporated to account for the blanketing effect.

In tuning the model inputs to reproduce an observed WD spectrum, it is desirable to achieve consistency for T$_{eff}$, log $g$ and abundance estimates that derive from 1) photometric fitting of the spectral energy distribution (SED), preferably using UV to optical and infrared wavelengths, 2) fitting the hydrogen and helium lines in profile and/or strength, and 3) demanding a balance in the abundances that come from different ionization states of the same element.   In principle, the model would meet the requirements of 1) through 3) while also achieving a reasonable fit of the broadening and symmetry of heavy element lines -- in practice, it can be challenging to reconcile all these points, but we find reasonable consistency for the WDs studied here.   Precise parameters are not required for our abundance analysis, since we are interested in the \emph{pollutant-to-pollutant} ratios, which are relatively insensitive to the particular model \citep[][also Section \ref{sec:abundances} of this paper]{klein2010}. 

In the analysis of helium dominated atmospheres with effective temperature below $\sim$16,000 K, we must deal with the difficulty that the parameters, T$_{eff}$ and gravity, $g$, are effectively coupled, such that T$_{eff}$/log $g$ fits to the helium lines are not unique.  This has led many authors (present authors included) to set log $g$ to the average value of 8.0, so that one may proceed with a fit to the temperature.  In the current analysis we took a different approach, similar to \cite{dufour2010}, who found that a larger than average gravity was required to fit simultaneously the features from two ionization states, Mg {\small I} and Mg {\small II}, in the extremely polluted helium-rich WD, SDSS J073842.56+183509.6 (hereafter, SDSS0738).   For each star in the present study we allow for a departure from the average gravity value, and instead work with a set of 3 T$_{eff}$/log $g$ solutions for T$_{eff}$ and log $g$ which do a reasonably good job of satisfying the spectral energy distribution (SED), He {\small I} line strengths, and the ionization balance between neutral and singly ionized species of Fe, Mg and Ca.     The model parameters used are listed in Table \ref{tab:modelsmass} and span a range of $\pm$300 or 400 K for effective temperature and $\pm$0.3 dex for log $g$.  
An additional constraint on the gravity could come from a sufficiently precise parallax measurement -- e.g. as \citet{kilic2008gd362} carried out for GD 362 -- and help narrow the range of possible log $g$ and T$_{eff}$ values.  To our knowledge, PG1225-079 and HS2253+8023 do not have reported parallaxes at this time\footnotemark[3]\footnotetext[3]{While G241-6 does have a parallax measurement \citep{vanAltena1995}, it is not precise enough to improve on the spectroscopically assumed T$_{eff}$/log $g$ used by \citet{zuckerman2010}.}.

\subsection{HS 2253+8023  T$_{eff}$/log $g$\label{sec:tefflogg_hs2253}}
The SEDs of our current models for HS2253 in the range T$_{eff}$=14000 - 14800 K, scaled to 2MASS photometry, are in agreement with the IUE data and GALEX photometry (Figure \ref{fig:IUE_fit}).  As \citet{limoges2010} pointed out, the strengths of some He {\small I} lines are especially gravity sensitive, and we make use of He {\small I} $\lambda$3820, $\lambda$4388, $\lambda$4713 and $\lambda$4922, in that regard.   We find the set of models: T$_{eff}$/log $g$ = 14000/8.1, 14400/8.4 and 14800/8.7, all give reasonable and roughly similar fits for most of the He {\small I} lines, shown plotted by the red line in Figure \ref{fig:hs2253_He}.  It turns out that the over-prediction of He {\small I} $\lambda$4713 mentioned by \citet{friedrich1999} is resolved by our use of higher than average gravity, although there remains an unaccounted-for shift of that line's center (top panel Figure \ref{fig:hs2253_He}).

Another reason we prefer models with a higher than average gravity, is we noticed that at the value log $g$=8.0, for a given element, even though we could get a reasonably good fit to the equivalent widths with the same abundance, for almost all lines the model profiles are too deep and too narrow;  this appeared to be the case for most elements.  Having worked with matching similar models to similar HIRES observations of similar WDs (GD40 and G241-6), we presume  that a mismatch in high-Z line broadening between the data and instrument-convolved model, is \emph{not} a problem of accounting for the instrumental broadening, but rather is indication that there is some intrinsic broadening needed in the model, such as a higher gravity.  We found that the models listed in Table \ref{tab:modelsmass} produced better overall fits to the high-Z line profiles than models of average gravity.  Similarly, the fit to the H$\alpha$ profile (Figure \ref{fig:hs2253_Halpha}) is better with higher gravity.  

Finally, with our choice models, the ionization balance for abundances of Mg and Fe naturally come out of the analysis of HS2253 at our desired consistency of within 0.1 dex between abundances from neutral and singly ionized states (Table \ref{tab:hs2253_abund}).  Obtaining  agreement in the ionization balance has proven challenging in other stars, including GD362, GD40 and PG1225 as described below.

\subsection{PG 1225-079  T$_{eff}$/log $g$\label{sec:tefflogg_pg1225}}
\begin{figure}[htbp]
\begin{center}
\includegraphics[width=110mm]{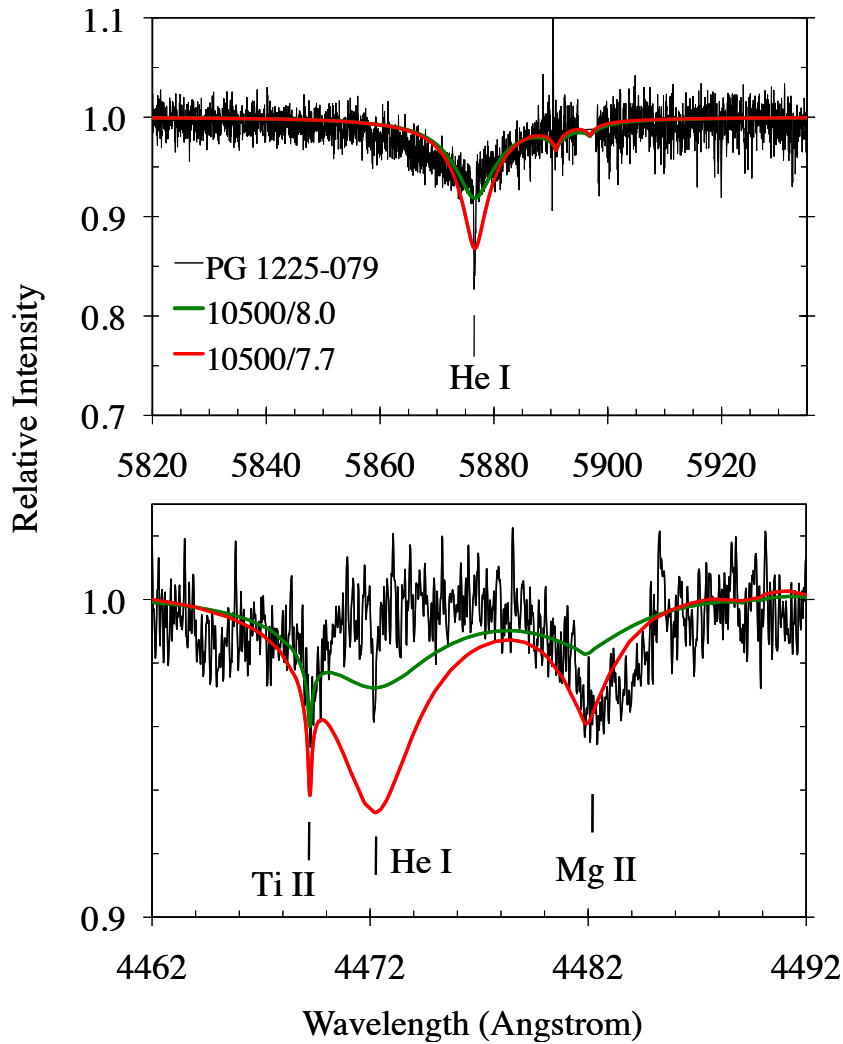}
\label{default}
\caption{Portions of the HIRES spectrum of PG1225-079 are plotted in black, displaying a helium line at He {\small I} $\lambda$5876 (top panel, data unsmoothed) and the region near He {\small I} $\lambda$4472 including the detection of Mg {\small II} $\lambda$4481 (lower panel, data smoothed by a 3 point boxcar average).  Wavelengths are in air and a heliocentric rest frame.   The upper panel spans a region across 2 echelle orders causing the gap near $\lambda$5896, which is simply filled in with a horizontal line.  As described in Section \ref{sec:tefflogg_pg1225}, it is challenging to simultaneously fit the He {\small I} lines and achieve an ionization balance for iron; the red curve is our best compromise.   The downward spike at the He line center is not noise, rather our current models do not predict the deep narrow core observed in this star's He {\small I} $\lambda$5876 line, see also Figure \ref{fig:3He5876cores} and the last paragraph of Section \ref{sec:obs_meas}.  The sodium upper limit reported in Table \ref{tab:pg1225_abund} is 0.15 dex higher than that of the synthetic Na {\small I} D line features near $\lambda$5890\AA\ shown here.  Lower Panel:  Instead of a broad shallow helium line, a weak narrow feature (EW $\sim$10 m\AA) appears, and is centered at He {\small I} $\lambda$4472 where no high-Z lines are expected.  Perhaps this is analogous to the narrow core of the He {\small I} 5876 line.  A weak narrow line also appears in the data at He{\small I} $\lambda$3889.  The 10500/8.0 model (green curve) shown for comparison was not used for abundance determinations so the Mg abundance of this model is not matched to the data.}
\label{fig:pg1225_He5876}
\end{center}
\end{figure}

\begin{figure}[htbp]
\begin{center}
  \includegraphics[width=140mm]{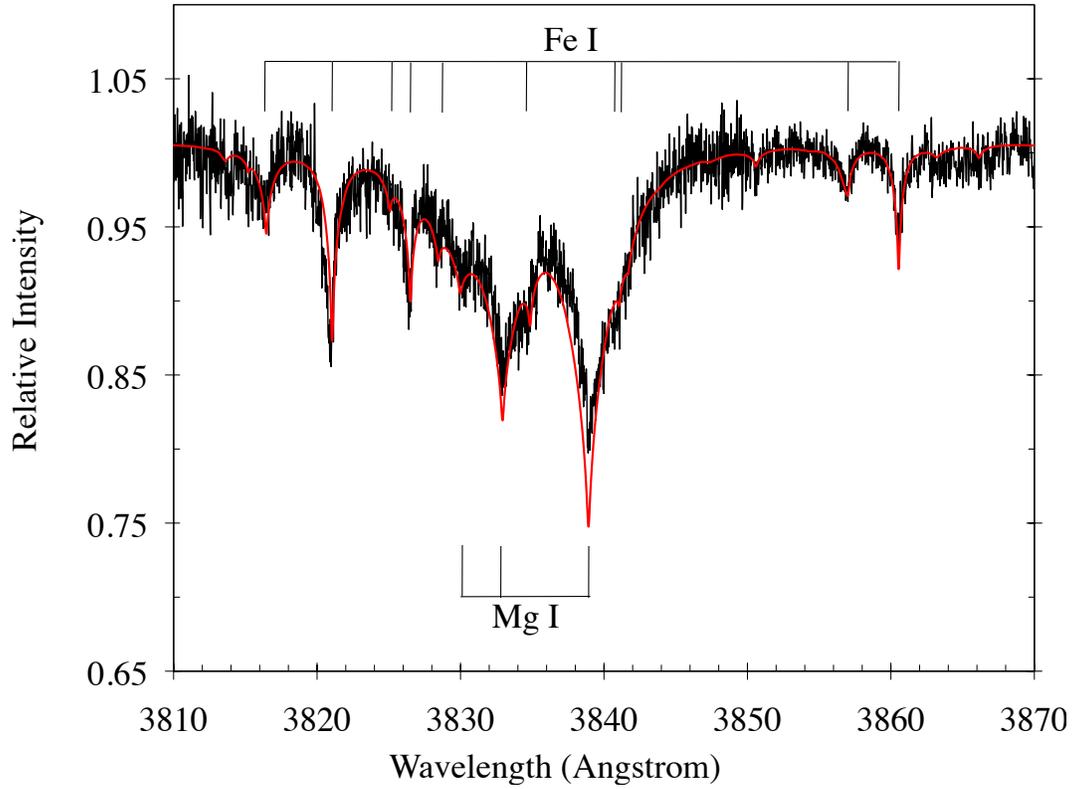}
\caption{Unsmoothed portion of the combined HIRES spectrum of PG1225-079 is plotted in black, displaying a Mg {\small I} triplet and many Fe {\small I} lines.  Wavelengths are in air and a heliocentric rest frame.  A T$_{eff}$=10,800 K log $g$ = 8.0 model, convolved with a Gaussian to approximate the instrumental resolution and with element abundances from Table \ref{tab:pg1225_abund}, is plotted in red.}
\label{fig:pg1225_MgI}
\end{center}
\end{figure}

\begin{figure}[htbp]
\begin{center}
  \includegraphics[width=140mm]{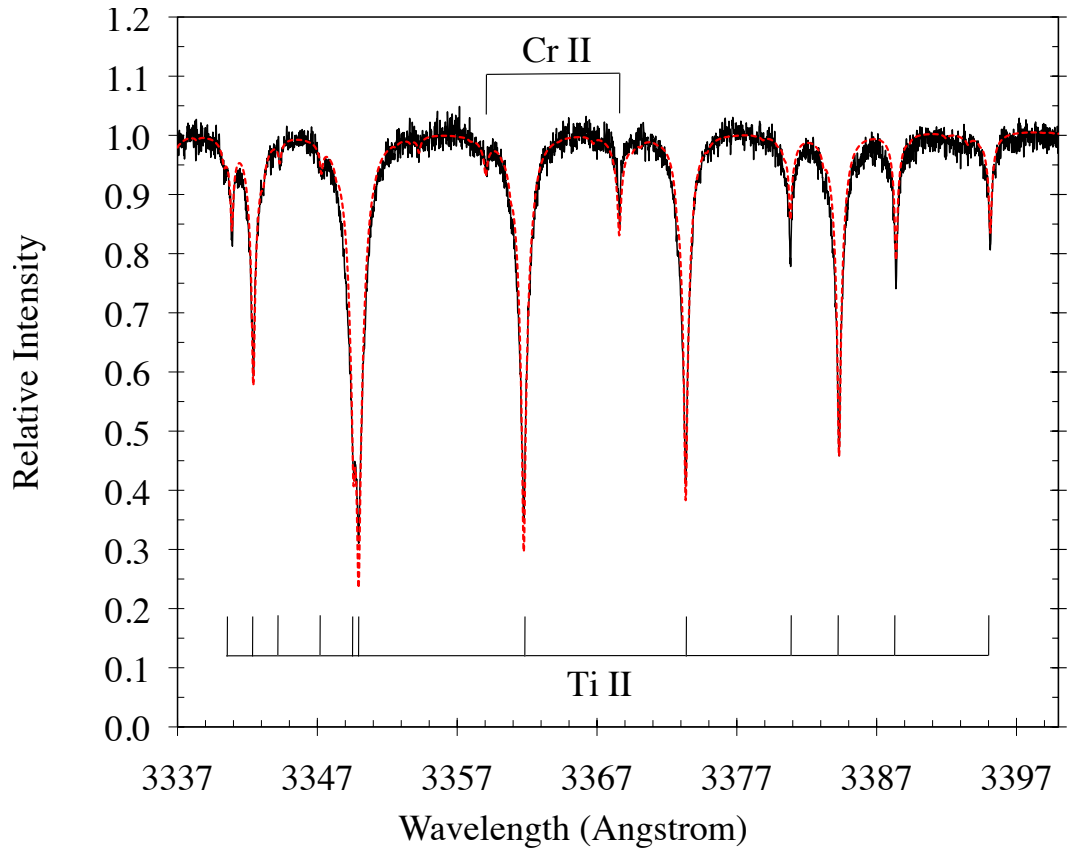}
\caption{PG1225-079.  Similar to Figure \ref{fig:pg1225_MgI}, but for Ti and Cr.}
\label{fig:pg1225_Ti}
\end{center}
\end{figure}

 At temperatures $<$12,000 K, the optical He {\small I} lines begin to disappear; He {\small I} $\lambda$5876 (Figure \ref{fig:pg1225_He5876}) is the strongest line observed.  In this range of T$_{eff}$, only a very small fraction of the photospheric helium is involved in the formation of optical helium lines, since they arise from levels of high excitation energy, e.g. 20.9 eV above ground.  Small changes in the atmosphere may have a large impact on the He line appearance, and we know that there is some unmodeled activity in this stars' upper atmosphere as is apparent from the emission cores of the Ca {\small II} H\&K lines (Figure \ref{fig:pg1225_CaHK}) and the narrow core of He {\small I} $\lambda$5876 (Figures \ref{fig:pg1225_He5876} \& \ref{fig:3He5876cores}).    For the discussion of atmospheric parameters that follows, we ignore the narrow core of the He {\small I} $\lambda$5876 feature described in Section \ref{sec:obs_meas}, and only consider the broad component of He {\small I} $\lambda$5876. 
 
The SEDs of our current models for PG1225, scaled to 2MASS photometry, are in good agreement with the IUE spectrum and GALEX photometry (upper panel Figure \ref{fig:IUE_fit}) for T$_{eff}$ near 10500 K, as used in previous work by \citet{wolff2002} with log $g$ = 8.0.  
 However, looking at the ionization balance, a 10500/8.0 model results in a 0.2 dex difference between iron abundances derived from Fe {\small I} and Fe {\small II}.   This is in the sense that Fe {\small I} lines are over-predicted and the discrepancy increases to 0.4 dex with a higher gravity model at 10500/8.3.  Inversely, we find at T$_{eff}$=10500 K, the two states of Fe are brought into marginal agreement (within 0.03 dex uncertainties -- Table \ref{tab:pg1225_abund}) by lowering the gravity to log $g$=7.7.  
 At this point in the T$_{eff}$/log$g$ parameter space, the 10500/7.7 model produces a He {\small I} $\lambda$5876 line with a predicted EW that matches the measured value at 1.5 \AA, but the predicted profile is somewhat narrower and deeper than observed (red curve in Figure \ref{fig:pg1225_He5876}).  Also, the He {\small I} $\lambda$4472 is not matched.   The model helium lines can be made shallower by increasing the gravity or decreasing the temperature, but always at the expense of degrading the ionization balance and the EW fit.  As was the case for HS2253, there is a degeneracy of model solutions that fit the equivalent width of the helium line -- for example 10500/7.7, 10800/8.0, and 11100/8.3 all produce a similar strength (and depth) He {\small I} 5876 feature.   Accepting the imperfections in the helium line profile fitting, we favor these models as our best compromise since they: 1) are in reasonable agreement with the UV/optical/IR spectral energy distribution, 2) fit the He {\small I} $\lambda$5876 equivalent width, and 3) support Fe {\small I}/Fe {\small II} abundance agreement to within 0.15 dex.  We use this set of three models for establishing pollutant-to-pollutant abundance ratios, which we show in Section \ref{sec:abundances}, are not very sensitive to the particular choice of model. 

\section{ABUNDANCES\label{sec:abundances}}   

Abundances are calculated for each line by linearly scaling the model input abundance by the ratio of EWs measured in the observed spectrum (Table \ref{tab:linelist}) to that of the model.  In cases of blended features or otherwise noisy regions, a visual inspection of the profile fit aids the determinations.   For each element we use the average of the set of calculated line abundances as the input for the next iteration of the model.   It takes a few iterations until the average converges.  Individual abundances are given equal weight in the final average element abundance.   The EW measurement uncertainties are propagated through to an uncertainty in the final average abundance.   We compare that to the mean deviation for the set of lines of an element, and assume the larger error.

A challenge to abundance accuracy can come from discrepancies in the atomic line data and broadening parameters \citep{vennes2011}.  However,  with multiple lines of each element observed, biases due to errors in the atomic data are minimized.  We typically find a low dispersion in derived abundances for elements with multiple line detections (ranging from 0.02 - 0.15 dex) -- generally comparable to, or smaller than, the propagated measurement uncertainties -- suggesting that the errors due to line data (including broadening parameters) are not a big factor in the current results. 

 Tables \ref{tab:hs2253_abund} and \ref{tab:pg1225_abund} list the pollutant  abundances relative to helium, derived from the different models considered for each star.  
  This explicitly shows that the choice of model temperature and gravity are a significant source of systematic uncertainty in absolute abundance determinations.    The absolute (pollutant-to-helium) abundances vary by 0.2-0.4 dex with the modest range of model changes we apply.  Also, estimates of the minimum accreted mass span a factor of ten, due to the fact that the calculated size of the convective zone has a strong dependence on the choice of model parameters as shown in Table \ref{tab:modelsmass}.   Fortunately for our goals, the relative (pollutant-to-pollutant) element abundances are rather homologous from one model to another with ratios varying only $\sim$40\% or less, a comparable contribution to the uncertainty as from the propagated measurement error.  The upshot is that the relative composition of the polluting elements can be measured to a greater precision than that of the absolute mass, or rate, of material accreted. The relative element abundances with respect to the fiducial -- magnesium -- are given in Table \ref{tab:abund_ratios}.

Comparisons of abundances with prior work is possible for H, Mg, Si, Ca and Fe, although previous uncertainties are typically large (0.2-0.7 dex), and tend to propagate to a sizable range for the abundance ratios.   For PG1225, the HIRES abundances relative to helium are nearly all within the uncertainties of those quoted by \citet{wolff2002}.  Within the uncertainty limits, pollutant-to-pollutant ratios agree between previous and present analyses.  From the HIRES data of PG1225, we obtain abundances for H, Mg, Ca, Sc, Ti, V, Cr, Mn, Fe, and Ni, and upper limits on O, Na, Al, and Sr, with an upper limit for Si that is consistent with its UV detection and analysis \citep{wolff2002}; an upper limit for carbon is taken from the UV work \citep{wolff2002}.
For HS2253, the nominal HIRES abundances with respect to helium are all lower than those quoted by \citet{friedrich1999} by at least 0.6 dex, most likely due to the differences in T$_{eff}$/log$g$ model inputs, particularly the much larger gravity used in the present work.  Nonetheless, the pollutant-to-pollutant abundances derived from HIRES data and \citet{friedrich1999} are mostly in agreement, with the exception of the Fe/Mg ratio.  Previously, HS2253 appeared iron-rich (by number, Fe/Mg = 2, with 1.2 to 3.3 covering the range of uncertainty);  the present HIRES analysis yields a ratio that is nominally half as large (Fe/Mg = 0.89 $\pm$ 0.18), and similar to Solar and Bulk Earth values. 
We obtain abundances in HS2253 from the HIRES data for H, O, Mg, Si, Ca, Ti, Cr, Mn, Fe, and a marginal detection of Ni;  we derive upper limits on Na, Al, Sc, V, and Sr, while an upper limit for carbon is taken from the UV work \citep{friedrich1999}.

\section{ACCRETION--DIFFUSION MODELS\label{sec:accretiondiffusion}}

As mentioned in Section \ref{sec:intro}, a helium-dominated polluted WD system has an evolving interplay of accretion and diffusion, since these WDs have relatively long settling times.  Calculated settling times specific to our particular models of HS2253 and PG1225 for the various elements studied are given in Tables \ref{tab:hs2253_abund} and \ref{tab:pg1225_abund}.  As described in \citet{koester2009}, the accretion--diffusion situation is roughly categorized by 3 phases: 1) an early phase in which accretion began ``recently'', that is, before differential settling has had time to significantly alter the abundance ratios -- the observed atmospheric abundances directly represent the accreted ones, 2) a steady state with ongoing accretion, where the observed abundances are related to the accreted ones by ratios of settling times, and 3) a dissipation phase after accretion has slowed, paused or ended, and elements with shorter settling times disappear faster than those with longer settling times.    We use these benchmarks to characterize our observations.

If accretion has been going on longer than a couple of settling times, then a steady state interpretation may be appropriate.  This assumes a semi-continuous flow of material lasting for $\sim$1 Myr or more in the systems under study here. 
To calculate the steady state abundances, for each element we take the average over different models of the ratios of settling times, and use the average ratios to apply a differential settling ``correction'' to the abundance ratios of Table \ref{tab:abund_ratios} as prescribed by Equation 7 of \citet{koester2009}.   It is worth noting that analogous to the abundances, while the settling times themselves vary significantly from one model to another, the pollutant-to-pollutant ratios of settling times vary very little (RMS of most $<$ 3\%, all $<$ 8\%) with the model variations for a given source.  For each element, the RMS's of diffusion data among models are propagated in quadrature into the steady state abundance uncertainties. 

 The principle result of applying a steady state ``correction'' is an increase in the abundance of iron and other heavy elements relative to the lighter elements in the inferred parent body.   This effect is apparent in Figure \ref{fig:cond_temp} where the early phase ($\equiv$ convective zone = CVZ) abundances are plotted as filled markers, while the the open markers represent parent body abundances for the system in a steady state.   For the set of elements plotted in Figure \ref{fig:cond_temp}, Mg and Si are the lightest and have the longest settling times. 
 Since Mg is the fiducial, applying a settling ``correction'' moves all the points of the plot up, relative to Mg, as seen in going from early phase to steady state values.  The difference between an early phase and a steady state interpretation is most pronounced in the O/Fe ratio, which can be seen by comparison of the bulk composition bar graphs, Figures \ref{fig:bar_graph_early} and \ref{fig:bar_graph_steady}.

\section{RESULTS AND DISCUSSION\label{sec:resultsdiscussion}}

From Table \ref{tab:abund_ratios} we find the accreted abundances are not in Solar proportions for the volatiles H, C, and O (HS2253), verifying that these systems are neither accreting from the interstellar medium nor gas giant planets.  Rather for example, the distinctively low carbon abundances in both HS2253 and GD40 are such that C/Mg is more than an order of magnitude below the Solar abundance and therefore more similar to Earth  \citep{jura2006}.  It is helpful to compare all measurements to a set of values such as the bulk Earth.  In the following we will discuss the abundance ratios of Table \ref{tab:abund_ratios}, mainly through Figure \ref{fig:cond_temp}, in which the observed abundance ratios are normalized to Mg and the bulk Earth, and Figures \ref{fig:bar_graph_early} and \ref{fig:bar_graph_steady} in which we compare the major element compositions of inferred parent bodies with the bulk Earth.   

HS2253 and PG1225 each have the high-Z mass of a large asteroid $\sim$10$^{22}$ g  currently in their convective zones (Table \ref{tab:modelsmass})  --  similar to GD362 \citep{zuckerman2007} and GD40 \citep{klein2010}.     HS2253 has one of the highest time-averaged mass accretion rates of all polluted white dwarfs, being comparable to GD362 \citep{jura2009apj,koester2009} at $\sim$10$^{10}$ g s$^{-1}$, with SDSS0738 apparently being the current champion at $\sim$10$^{11}$ g s$^{-1}$ based on \citet{dufour2010} and our settling time calculations parallel to \citet{koester2009}.  The time-averaged mass accretion rate of rocky material in PG1225 is a few $\times 10^8$ g s$^{-1}$, about a factor of 5-10 less than GD40.

\subsection{HS 2253+8023  Composition\label{sec:discussion_hs2253}}

If a WD accretes a rocky planetary body, then the relative element abundances will have the distinctive proportions that arise from mineral oxides \citep{jura2009apj, klein2010}, made up of MgO, Al$_2$O$_3$, SiO$_2$, CaO, TiO, Cr$_2$O$_3$, etc.\ with iron coming from either FeO, Fe$_2$O$_3$, or Fe metal if a differentiated core formed.  Another possible vehicle for oxygen in a terrestrial-type planetary body is water ice, which depends on how much hydrogen is present.
 For HS2253, the HIRES measurement of oxygen plus all major elements, some minor ones, and hydrogen, allows us to evaluate the balance of oxides and water in this system according to Equations 2 and 3 of  \citet{klein2010}.

Looking at the oxide balance in an early phase interpretation for HS2253, we find that the number ratio of oxygen that can be accounted for by oxides with the other elements (using half the iron in FeO, half in Fe$_2$O$_3$, plus aluminum at an Earth-like ratio of Al/Mg = 0.086) is O/Mg = 3.8 $\pm$ 0.6.   Since the observed amount of oxygen is O/Mg = 5.4 $\pm$ 1.1, at the limits of the uncertainties  the oxygen can be carried in rocky minerals if iron is predominantly in oxide form;  water is not needed to balance the oxides.   Generally though, due to the ambiguity of the state of iron and since iron is a major element, there is a sensitive trade-off between metallic iron and water in balancing the oxides.   In HS2253 the early phase abundances can be attributed solely to mineral oxides, but could involve up to  $\sim$15\% water by mass (limited by the observed hydrogen abundance).

   In the steady state for HS2253, the number ratio of oxygen expected to be associated with the observed major elements, Mg, Al, Si, Ca, and Fe, as minerals (half the iron in FeO, half in Fe$_2$O$_3$, plus an Earth-like ratio for Al)  is O/Mg = 4.7 $\pm$ 0.8, and the ratio of of observed oxygen coming from the parent body is O/Mg = 4.5 $\pm$ 0.9.   In the steady state the ratios agree very well with a rocky mineral oxide composition.   Again, since some Fe may have been metallic and there is a range of uncertainty overlap in the O/Mg ratios, there is still sufficient oxygen to allow for some water;   the amount of possible water in the parent body under a steady state interpretation is by mass $\lesssim$7\%, limited by the 
observed hydrogen abundance.   While HS2253 does not display an infrared excess, it can be seen from Figures \ref{fig:bar_graph_early} and \ref{fig:bar_graph_steady} that this heavily polluted WD has atmospheric abundance ratios that are similar  to GD40 -- an analogous WD that does display an infrared excess.

Could the system be in a dissipating phase?  Not likely, since the oxides and water are well balanced in both the early phase and steady state interpretations; a dissipative interpretation will eventually run into difficulty with oxides.  From the lower panel of Figure \ref{fig:cond_temp} in an early phase interpretation, the abundances of Ti, Mn, Cr, Fe, and Ni, are in excellent agreement with bulk Earth values (the Ca overabundance and Si underabundance are similar to GD40).   As accretion--diffusion proceeds, the ratios move up to the steady state values.      If dissipation takes over, the inferred abundance ratios would be driven even further up compared to Mg and Si, while O/Mg decreases, and would imply a parent body that is somehow depleted of the major elements Mg, Si, and O, compared to high abundance ratios of Ca, Ti, Mn, Cr, Fe and Ni.  We have no physical explanation for how that may occur.    Thus, we prefer the simpler explanation that HS2253 most likely accreted material with abundances predominantly similar to bulk Earth and is somewhere at or between an early phase and a steady state of accretion and diffusion.

\subsection{PG 1225-079 Composition\label{sec:discussion_pg1225}}

In addition to significant high-Z element pollution, PG1225 has a considerable amount of hydrogen in its convective zone.  Where did it come from?  For GD 362, \citet{jura2009apj} showed that the presence of circumstellar dust, high-Z material and copious hydrogen can simultaneously be explained by the accretion of a large ice-rich planet.   Figure 1 of \citet{jura2009apj} illustrates that the atmospheric pollution by hydrogen in GD362 and GD16 is anomalously high compared to other known helium-dominated externally polluted WDs\footnotemark[4]\footnotetext[4]{See also \citet{tremblay2011} for 2 more recently uncovered candidates of this unusual type of WD --  helium-dominated with large quantities of atmospheric hydrogen.} of their evolutionary epoch, 10,500 K $<$ T$_{eff}$ $<$ 11,500 K, i.e.\ cooling age 0.4 - 1 Gyr.  At a similar cooling age, our set of possible models for PG1225  indicate a hydrogen mass which can range from 6$\times$10$^{22}$ to 8$\times$10$^{23}$ grams.  Referring to Figure 1 of \citet{jura2009apj} this is similar, perhaps on the high-end, but not extraordinary, compared to the average amount of accreted hydrogen in the set of helium-dominated WDs with detected hydrogen at this age.  \citet{juraxu2010} suggest that the origin of the hydrogen in these more ordinary WDs may be from the accretion of ice-rich planet(esimal)s which survive post-main sequence evolution.

Unfortunately, we cannot evaluate a water and oxide balance for PG1225 since the O{\small I} $\lambda$7775 lines are not visible in its HIRES spectrum.   If the detected elements came from a rocky parent body, then the amount of oxygen (by number) associated with mineral oxides is O/Mg $\sim$ 5, which is far below the sensitivity of our observations that yield an upper limit of O/Mg $<$ 60 (Table \ref{tab:abund_ratios}).   With just a small portion of the allowed O/Mg number ratio allocated to rocky oxides, the remaining number ratio that could be associated with water is O/Mg $<$ 55, but that only accounts for less than 10\% of the observed hydrogen.   In order to explain all of the observed hydrogen as being derived from icy parent bodies, either an ongoing steady state of atmospheric settling from a somewhat large parent body (perhaps a Ceres analog) or previous ice-rich accretion event(s), would be implied.

PG1225's spectrum is very rich with high-Z lines and the numerous newly detected elements display an interesting abundance pattern.  In the top panel of Figure \ref{fig:cond_temp}, Mg, Cr, Mn, Fe and Ni, have abundance ratios similar to bulk Earth values, but we find that the refractory elements -- Ca, Sc, Ti and V -- are all a factor of $\sim$2--3 higher abundance than in the Earth.  High Ca and Ti abundance ratios have previously been  measured in GD362 \citep{zuckerman2007} and  GD40 \citep{klein2010}.  As shown in the bottom panel of Figure \ref{fig:cond_temp}, HS2253 also has a somewhat enhanced Ca/Mg ratio.  The enhancement of refractory element abundances with condensation temperatures $>$1400K is significant and uniform in PG1225, and the upper limits derived for Al and Sr are consistent with the trend.   One possible explanation for this observation is that the material accreted onto PG 1225-079 derived from a planetary body that formed in a higher temperature environment than did the Earth.  The simulations of \citet{bond2010} suggest that refractory-rich exoplanets should be a common outcome of planet formation in nebular regions sufficiently close to the host star.

PG1225 has just a subtle infrared excess appearing only at 7.9$\micron$ \citep{farihi2010, kilic2008dust}.  \citet{farihi2010} raise the possibility that it could be due to a relatively narrow dust ring with an inner hole cleared by accretion, suggesting that the system is in a dissipative phase.  Referring to Figure \ref{fig:cond_temp}, since all the refractory elements have shorter settling times than Mg, any ``correction'' for settling causes an even greater overabundance of refractory species inferred in the parent body.  In other words, contrary to the expectation from a dissipative phase, the heavy elements do not appear sunk.  However, the settling times in this WD are quite long $\sim$Myr, and disk lifetimes may be shorter; a large range has been suggested $\sim$10$^5$--10$^9$ yr \citep[][and references therein]{jura2009apj}.
Thus it is conceivable that we could be viewing the system at a point toward the end of an accretion episode but before dissipative differential settling has a significant effect on the atmospheric ratios.  Whatever the case regarding circumstellar dust, the atmospheric element ratios imply that PG1225 most likely accreted material with abundances predominantly similar to bulk Earth, but enhanced in refractory species, and the system is somewhere in or near an early phase or steady state of accretion and diffusion.

\subsection{Helium-Rich Highly-Polluted White Dwarfs\label{sec:the_set}}
Finally we discuss the set of highly polluted helium-dominated white dwarfs that have been comprehensively studied.  To date,  5 helium-dominated WDs have $\geq$8 elements reported (GD362, GD40, G241-6, HS2253+8023, PG1225-079) plus 2 more (SDSS0738, GD61) have measurements of all major (O, Mg, Si, and Fe) terrestrial planetary constituents. 

Though our focus in this paper is on compositions in heavily polluted, helium-dominated WDs, we recognize the importance of the study of heavily polluted, hydrogen-dominated WDs, such as NLTT43806 \citep{zuckerman2011} and GALEX J193156.8 + 011745 \citep{vennes2010}.  With the detection of circumstellar dust \citep{debes2011,melis2011arXiv} and all the major element species (O, Mg, Si, and Fe), GALEX J193156.8 + 011745 (GALEX1931) is almost certainly accreting a planetary body; with settling times on the order of weeks, accretion must be happening now.  However we note there is currently a troubling inconsistency between the modeled results of different groups \citep{vennes2011, melis2011arXiv}, which is due in part to a large, unexplained difference for the assumed effective temperature ($\Delta$T$_{eff}$ $>$ 2,000 K) of GALEX1931, and in part to the treatment of atmospheric settling, which for warm (T$_{eff}$ $\sim$ 20,000 K) hydrogen-dominated WDs, is complicated by the lack of a convective envelope.

Figures \ref{fig:bar_graph_early} and \ref{fig:bar_graph_steady} represent the compositions by mass of major elements in the helium-dominated polluted WDs in which oxygen has been detected, or a meaningful upper limit set (GD362).  Lacking an oxygen detection, PG1225 is not on the bar graphs.   There are two different bar graphs showing two different interpretations of the parent body abundances, in the early phase and in the steady state.  The graphs are related by ratios of settling times as described in Section \ref{sec:accretiondiffusion}.     We are interested in comparing the bulk accreted compositions, and since hydrogen never settles out of the convective zone, we cannot be certain how much of it is associated with the current heavy element pollution.  We do not include hydrogen in Figures \ref{fig:bar_graph_early} and \ref{fig:bar_graph_steady}, which has a negligible effect on the graph appearance for bulk Earth, HS2253, GD40, SDSS0738 and G241-6, being less than 1\% of the accreted mass in those sources.  On the other hand, in GD362 hydrogen is over 99\% of the accreted mass and in GD61 it is over 75\%.   As discussed in the preceding section, scenarios have been proposed to explain the presence of hydrogen from the accretion of ice-rich planetary bodies.  Now we consider the compositional possibilities from the perspective of oxygen.  

In each of the WD systems with detected oxygen, there is always enough O to account for the delivery of the major\footnotemark[5]\footnotetext[5]{Aluminum may also be expected in comparable mass as Ca, but it is more difficult to detect.  Within uncertainties, the undetected presence of Al at an Earth-like abundance will not substantially change the oxide balance.} elements Mg, Si, Ca, and Fe, in the form of a terrestrial-type planetary body.   
The upper limit of oxygen in GD362 is just enough to have come from Mg, Si, Ca and Fe in oxides, with the possibility of metallic iron.  Since there is very little ``spare'' oxygen in GD362, in order to account for the hydrogen, high-Z elements, infrared excess, and lack of x-ray luminosity in a unified picture, \citet{jura2009apj} calculate a scenario in which settling and accretion have been ongoing for a long time, implying the consumption of a large ice-rich planetary body at least the size of Callisto.

 On the right side of the bar graphs, the abundance ratios of GD61 imply the accretion of more oxygen than can be carried in rock.  An O ``excess'' can be the result of dissipative gravitational settling, which would make the abundances seem oxygen-rich and iron-poor.    Another way to get an O excess is by the accretion of an ice-rich parent body, as considered in detail by \citet{juraxu2010}.   The especially high O/Fe  and Mg/Fe abundance ratios in the atmosphere of GD61 look suspiciously similar to what is expected in a system where accretion has stopped or paused -- the heavier elements sink faster than the lighter ones.  However, along with measurements of all the major element abundances, \citet{farihi2011gd61} report an infrared excess for GD61.  Since the WD has plenty of hydrogen in its atmosphere, an O excess, and detected circumstellar material, \citet{farihi2011gd61} raise again the possibility of the accretion of an ice-rich planetary body. Due to abundance measurement uncertainties and the uncertainty of the accretion-diffusion phase, the results are yet inconclusive regarding the composition of the parent body accreted onto GD61.

G241-6 also appears to be a case with an O excess, even at the limits of the abundance uncertainties.  The upper limit on hydrogen in the atmosphere amounts to less than 0.6\% of the mass of detected contaminants, and therefore cannot sufficiently account for the excess oxygen by an origin in ice; the parent body must have been relatively dry.  The abundance pattern of G241-6 is similar to GD40 but with Mg, Si, Ca and Fe appearing ``sunk'' (compare the GD40 early phase bar of Figure \ref{fig:bar_graph_early} with G241-6's steady state bar of Figure \ref{fig:bar_graph_steady}).  In other words, starting from GD40 abundances, if we could turn accretion off, then eventually gravitational settling will evolve the abundances into ratios similar to what we observe for G241-6.   Since we now know that G241-6 does not have an infrared excess (Xu \& Jura, \emph{Spitzer} data, in preparation), it is likely that G241-6 is in a dissipative phase, having originally accreted an ice-free planetary body with composition similar to GD40 and the bulk Earth.

The abundances in the remaining three: HS2253, GD40, SDSS0738, are all consistent with the major elements being derived from rocky mineral oxides, and a composition similar to bulk Earth being composed more than 85\% by mass in O, Mg, Si, and Fe, for either a steady state and/or early phase interpretation.  Overall their parent body compositions are depleted of volatiles;  in addition to the dearth of carbon in GD40 and HS2253, all of them have distinctively low abundances of hydrogen (compared to high-Z elements) and little, if any, excess oxygen -- again indicating a dearth of ice in the accreted material.   
 
While GD362 and GD61 are candidates for exoplanetary systems having accreted a substantial amount of water, the other four stars with oxygen measurements must have accreted relatively dry planetary bodies ($\lesssim$10\% H$_2$O by mass).  Furthermore, in all four of those cases, the pollution is most plausibly dominated by planet(esimal)s that formed with little or no ice, for reasons as follows.  GD40 and SDSS0738 both have an infrared excess, and as discussed by Jura (2008), the existence of a dust disk is best understood as resulting from the disruption of a single parent body, rather than multiple parent bodies -- which, arriving on trajectories of varied inclinations, tend to result in gaseous disks due to dust grain destruction by sputtering in high-velocity impacts.  With a minimum accreted mass of $\sim$10$^{22}$g, and a rock density of 3 g cm$^{-3}$, the radius of a single polluting parent body is greater than $\sim$100 km.   Thermal erosion calculations by \citet{juraxu2010} demonstrate that asteroids of radii $>$ 100 km would have retained at least 50\% of their internal ice, if present, through red giant evolution.  Thus the parent bodies polluting GD40 and SDSS0738 most likely formed without much ice.  

Could HS2253 and G241-6 be polluted by many small (radius $<$ 20-30 km), initially ice-rich asteroids whose internal water was lost during the host star's Asymptotic Giant Branch (AGB) phase?   Consider a model for HS2253 involving 25 km asteroids that were 25\% water by mass before the star was on the AGB.    According to Figure 4 of \citet{juraxu2010}, unless they originally orbited at less than 5 AU, 25 km asteroids would retain at least 10\% of their water during AGB evolution of a 3 $\Msun$ star.  At a density of 2.1 g cm$^{-3}$ (analogous to Ceres), the pre-AGB asteroid might have had 10$^{20}$ g of high-Z material, and 4 $\times$ 10$^{18}$ g of hydrogen (H) from ice.  After the AGB, the asteroid has 10$^{20}$ g of high-Z material, and 4 $\times$ 10$^{17}$ g of H.  With the middle model from Table \ref{tab:hs2253_abund}, the convective zone of HS2253 has $\sim$2 $\times$ 10$^{22}$ g of high-Z matter.  Therefore, it needs to have accreted about 200 asteroids during the most recent 10$^5$ years, or about 1 asteroid every 500 years.  During this time, the white dwarf would acquire 8 $\times$ 10$^{19}$ g of H, about a quarter of the total H ($\sim$3 $\times$ 10$^{20}$ g with the middle model from Table \ref{tab:hs2253_abund}), which the WD has accumulated over its entire cooling age of approximately 400 Myr.  The relatively small amount of H in the convective zone implies that no more than $\sim$600 similar asteroids could have arrived in the previous 400 Myr.  Why would the asteroids arrive at a rate of 0.002 yr$^{-1}$ recently and 2 $\times$ 10$^{-6}$ yr$^{-1}$ previously?  The result of this model exercise is analogous, but more extreme, for G241-6.  Another difficulty with the above hypothesis is that, even if there is a burst of impacts, one imagines that there would be a size distribution of impactors and, at least in the usual kind of model, most of the mass would be carried in the larger objects.  All that is required to produce the observed hydrogen is the accretion of single ice-rich asteroid of radius 100 km.  Unless the size distribution of asteroids is very steep, there could easily be one big asteroid for every 200 small ones.  Therefore, to explain the pollutions of HS2253 and G241-6 by swarms of initially ice-rich small asteroids, would require both a spike in time of number frequency and and a spike in the size distribution of the asteroids.  This double constraint seems implausible.
It is most likely that the planet(esimal)s which pollute these four now comprehensively studied WDs  -- HS2253, GD40, SDSS0738 and G241-6 -- were formed with little ice, and we are measuring the distilled elemental compositions of bodies which condensed out of protoplanetary disks at radii interior to any snow line.

\section{CONCLUSIONS\label{sec:conclusions}}

   We have obtained and analyzed high-resolution optical spectra of HS2253+8023 and PG1225-079 with HIRES at Keck Observatory.  This work adds two more systems to the small but fast-growing sample of polluted white dwarfs with detailed and comprehensive analyses. 
   Four out of six objects with oxygen measurements most likely accreted planet(esimal)s of mineral oxide constitution, which were nearly ice-free, consistent with formation interior to a snow line.    It is clear that extrasolar planetary systems produce rocky bodies that are compositionally similar to terrestrial planets in our own solar system; Earth-like planet(esimal)s apparently do form elsewhere in the galaxy.

\section{ACKNOWLEDGEMENTS}
The authors would like to thank S. Xu and C. Melis for their contributions, and an anonymous referee for suggestions that improved the resulting manuscript.  This work has been supported by grants from the NSF and NASA to UCLA.  We acknowledge extensive use of VALD, the Vienna Atomic Line Database \citep{piskunov1995, ryabchikova1997, kupka1999, kupka2000}.  This publication has made use of NASAÕs Astrophysics Data System.
We thank the Keck Observatory staff for their support.  The data presented herein were obtained at the W.M. Keck Observatory, which is operated as a scientific partnership among the California Institute of Technology, the University of California and the National Aeronautics and Space Administration. The Observatory was made possible by the generous financial support of the W.M. Keck Foundation.  We recognize and acknowledge the very significant cultural role and reverence that the summit of Mauna Kea has always had within the indigenous Hawaiian community.  We are most fortunate to have the opportunity to conduct observations from this mountain. 

\bibliographystyle{apj}
\bibliography{apj-jour,myrefs}

\newpage

\begin{table}[htdp]

\caption{Keck/HIRES Observations}
\begin{center}
\begin{tabular}{lccccccc}
\hline 
\hline

WD & UT Date & Approx. Seeing  & Collimator & $\lambda$ range & Resolution  & Exposure   \\ 
 & &  (visual) &  &   (\AA) &   ${\lambda}\over{\Delta\lambda}$  & (seconds)  \\ 
\hline
G 241-6  & 2007 Nov 20 & 0\farcs8 & Blue & 3120 -- 5950 & 40000 & 3000 \\
   & 2008 Aug 06 & 1\farcs3 & Blue & 3140 -- 5950 & 40000 & 2 x 1500  \\ 
   & 2008 Nov 14 & 1\farcs1 & Red & 4500 -- 9000   & 40000 & 1800+2000 \\ 
HS 2253+8023 & 2008 Aug 06 & 1\farcs3 & Blue & 3140 -- 5950 & 40000 & 1800+2400  \\ 
   & 2008 Nov  14 & 1\farcs1 & Red & 4500 -- 9000  & 40000 &   2 x 2000  \\ 
   & 2008 Nov 15 & 1\farcs4 & Red & 4500 -- 9000 & 28000  &  2 x 1800 \\
PG 1225-079 & 2008 Feb 13 & 0\farcs8 & Blue & 3120 -- 5950 & 40000 & 2 x 1600 \\
    & 2008 Feb 14 & not avail & Blue & 3120 -- 5950 & 40000 & 2600 \\
   & 2008 Feb 26 & 0\farcs8 & Red & 4600 -- 9100 & 40000 & 4000 \\
\hline

\end{tabular}
\label{tab:obs}
\end{center}
On 2008 Feb 14 the sky was partly cloudy with an estimated visual extinction of $\sim$0.5 mag.  The rest of the  observations occurred during mostly clear skies.  Gaps in blue coverage with our setups occur between 4010-4100 \AA, and 5990-6100 \AA.  There are many small gaps in red coverage.
\end{table}

\begin{center}
\begin{longtable}{ l l c c c c c c} 
\caption{Absorption Line Measurements of HIRES Spectra} \\ 
\hline
\hline

Ion & $\lambda$$^a$ & $\chi$ & log $gf$  &  \multicolumn{3}{c}{ Equivalent Width (m\AA)}  \\
	&(\AA) 	&   (eV)    &  &	PG 1225-079	&	HS 2253-8023	&	G241-6 $^c$	\\

\hline
\endfirsthead

\multicolumn{6}{c}{Table 2 --- \emph{Continued}} \\ 
\hline
\hline
Ion & $\lambda$$^a$ & $\chi$ & log $gf$  &  \multicolumn{3}{c}{ Equivalent Width (m\AA)}  \\
	& (\AA) &  (eV) &  &	PG 1225-079	&	HS 2253-8023	&	G241-6 $^c$	\\
\hline
\endhead

  \multicolumn{3}{l}{{Continued on Next Page\ldots}} \\
\endfoot
  
 \hline
\endlastfoot

H {\small I} & 6562.79 & 10.20	& 0.71	&	10,500 (2,000)	&	850 (120)	&    $<$ 250 \\ 
H {\small I} & 4861.32 &  10.20 & -0.02	&	3,500 (1,500)  &  120 (30)   &  -- \\ 
   O {\small I} &   7771.94 &   9.15 &   0.37 &   $<$ 250  &   350 (90) &   250 (50) \\ 
   O {\small I} &   7774.16 &   9.15 &   0.22 &    --  &   190 (50) &   170 (50) \\ 
   O {\small I} &   7775.39 &   9.15 &   0.00 &    --  &   220 (110) &    90 (40) \\ 
   O {\small I} &   8446.35 &   9.52 &   0.24 &    --  &   380 (190) &   210 (50) \\ 
   Na	{\small I} & 5889.95 &   0   & 0.12  &	$<$ 50 &  $<$ 120 &  --  \\
  Mg {\small I} &   3829.35 &   2.71 &  -0.23 &   in 3800 interval &    50 (6) &    15 (5) \\ 
  Mg {\small I} &   3832.30 &   2.71 &  -0.36, 0.12 & in 3800 interval  & 220 (40) & 75 (14) \\ 
  Mg {\small I} &   3838.29 &   2.72 &   0.39, -0.35 & in 3800 interval & 470 (40) & 106 (14) \\ 
  Mg {\small I} &   5172.68 &   2.71 &  -0.45 &    30 (15) &    70 (40) &    31 (15) \\ 
  Mg {\small I} &   5183.60 &   2.72 &  -0.18 &    90 (45) &   170 (40) &    52 (11) \\ 
  Mg {\small II} & 4481 &  8.86 &  0.74,-0.56,0.59 & 90 (20) & 1040 (110) & 470 (50) \\ 
  Mg {\small II} &   7877.05 &  10.00 &   0.39 &    --  &   270 (110) &   110 (25) \\ 
  Mg {\small II} &   7896.04 &  10.00 &  -0.31 &    --  & in Mg{\small II} 7896.36 & in Mg{\small II} 7896.36 \\ 
  Mg {\small II} &   7896.36 &  10.00 &  0.65 &    --  &   460 (130)$^{b - Mg{\small II}}$ &   250 (40)$^{b - Mg{\small II}}$ \\ 
  Al {\small I}   &  3961.52	&  0.014 & -0.32  &  $<$ 40  &  $<$ 60 &  $<$ 30  \\
  Si {\small I}  &  3905.52	&	1.91	& -0.74	&	$<$ 35  &  -- & -- \\
  Si {\small II} &   3856.02 &   6.86 &  -0.41 &    --  &   190 (20) &    74 (19) \\ 
  Si {\small II} &   3862.59 &   6.86 &  -0.76 &    --  &   119 (13) &    38 (8) \\ 
  Si {\small II} &   4128.05 &   9.84 &   0.36 &    --  &    67 (25) &    27 (4) \\ 
  Si {\small II} &   4130.89 &   9.84 &   0.55 &    --  &   110 (17) &    35 (8) \\ 
  Si {\small II} &   5055.98 &  10.07 &   0.52 &    --  &   126 (27) &    53 (15) \\ 
  Si {\small II} &   6347.11 &   8.12 &   0.15 &    --  &   220 (70) &   113 (22) \\ 
  Si {\small II} &   6371.37 &   8.12 &  -0.08 &    --  &   160 (20) &    43 (16) \\ 
  Ca {\small I} &   4226.73 &   0 &   0.27 &   210 (60) &    --  &    --  \\ 
  Ca {\small II} &   3158.87 &   3.12 &  0.24 &  2600 (520) &   890 (70) &   200 (20) \\ 
  Ca {\small II} &   3179.33 &   3.15 &  0.50 & 3600 (720)$^{b - Ca{\small II}+Cr{\small II}}$  & 1180 (150)$^{b - Ca{\small II}+Cr{\small II}}$ &   250 (30) \\ 
  Ca {\small II} &   3181.28 &   3.15 &  -0.46 &    in Ca{\small II} 3179.33 & in Ca{\small II} 3179.33 & 47 (18) \\ 
  Ca {\small II} &   3706.02 &   3.12 &  -0.48 &  1100 (220) &   240 (40) &    45 (5) \\ 
  Ca {\small II} &   3736.90 &   3.15 &  -0.17 &  1700 (340)$^{b - Fe{\small I}}$  &   590 (50) &   113 (9) \\ 
  Ca {\small II} &   3933.66 &   0 &   0.10 &   39,000 $^d$ &  4700 (500) &  1300 (150) \\ 
  Ca {\small II} &   3968.47 &   0 &  -0.20 &   in Ca{\small II} 3933.66 &  2740 (360) &   790 (80) \\ 
  Ca {\small II} &   8498.02 &   1.69 &  -1.42 &  1350 (400) &   570 (80) &    --  \\ 
  Ca {\small II} &   8542.09 &   1.70 &  -0.46 &  5000 (1500) &  2340 (320) &   780 (90) \\ 
  Ca {\small II} &   8662.14 &   1.69 &  -0.72 &  4300 (1300) &  1720 (160) &   290 (90) \\ 
  Sc {\small II} &   3572.53 &   0.02 &   0.27 &    18 (5) &    --  &    --  \\ 
  Sc {\small II} &   3613.83 &   0.02 &   0.42 &    27 (8) &    $<$ 80  &    --  \\ 
  Sc {\small II} &   3630.74 &   0.008 &   0.22 &    in Fe{\small I} 3631.46 &    --  &    --  \\ 
  Ti {\small II}  &   3148.04  &  0	&	-1.22 & 29 (8) & -- & -- \\
  Ti {\small II} &   3154.19 &   0.11 &  -1.15 &    in FeII 3154.20  &  --  &  --  \\ 
  Ti {\small II} &  3161.20	& 0.11  & -0.69  &  20 (4) & -- & -- \\
  Ti {\small II} &   3161.77 &   0.12 &  -0.55 &    41 (8) &    --  &    --  \\ 
  Ti {\small II} &   3162.57 &   0.14 &  -0.38 &    54 (11) &    --  &    --  \\ 
  Ti {\small II} &   3168.52 &   0.15 &  -0.20 &    190 (30)$^{b - Fe{\small II}}$  &  25 (10) &    --  \\  
  Ti {\small II} &   3190.87 &   1.08 &   0.23 &   136 (27) &    37 (18) &    --  \\ 
  Ti {\small II} &   3202.53 &   1.08 &   0.07 &    74 (15) &    --  &    --  \\ 
  Ti {\small II} &   3217.05 &   0.03 &  -0.49 &   120 (24) &    40 (20) &    --  \\ 
  Ti {\small II} &   3218.26 &   1.57 &   0.05 &    47 (9) &    --  &    --  \\ 
  Ti {\small II} &   3222.84 &   0.012 &  -0.42 &   130 (26) &    43 (13)  &    --  \\ 
  Ti {\small II} &   3224.24 &   1.58 &   0.05 &    23 (5) &    --  &    --  \\ 
  Ti {\small II} &   3228.60 &   1.08 &  -0.23 &    in 3226 interval &    --  &    --  \\ 
  Ti {\small II} &   3229.19 &   0 &  -0.56 &   130 (50) &    36 (11)  &    --  \\ 
  Ti {\small II} &   3229.42 &   1.13 &  -0.11 &  in 3226 interval  &    --  &    --  \\ 
  Ti {\small II} &   3232.28 &   1.12 &  -0.22 &    20 (4) &    --  &    --  \\ 
  Ti {\small II} &   3234.51 &   0.05 &   0.43 &   620 (120) &   140 (30) &    34 (7) \\ 
  Ti {\small II} &   3236.57 &   0.03 &   0.24 &   520 (100) &    87 (20) &    34 (6) \\ 
  Ti {\small II} &   3239.04 &   0.012 &   0.07 &   430 (90) &    74 (22) &    21 (7) \\ 
  Ti {\small II} &   3239.66 &  1.08  &  -0.20 &  50 (15)  & -- & -- \\
  Ti {\small II} &   3241.98 &   0 &   -0.03    &  280 (60)  & 79 (12)    &    --  \\ 
  Ti {\small II} &   3248.60 &   1.24 &   0.37 &   134 (27) &    65 (35) &    --  \\ 
  Ti {\small II} &   3251.91 &   0.012 &  -0.59 &    91 (18) &    --  &    --  \\ 
  Ti {\small II} &   3252.91 &   0.03&  -0.48 & 150 (30)$^{b - Ti{\small II}}$ &  52 (10)$^{b - Ti{\small II}}$  &  --  \\ 
  Ti {\small II} &    3252.94 &   1.08 &   -0.25 &   in Ti{\small II} 3252.91 &   in Ti{\small II} 3252.91  &    --  \\ 
  Ti {\small II} &   3254.25 &   0.05 &  -0.56 &    96 (19) &    40 (16) &    --  \\ 
  Ti {\small II} &   3261.58 &   1.89 &  0.53 &   190 (40)$^{b - Ti{\small II}}$ &  65 (29)$^{b - Ti{\small II}}$  &    --  \\ 
  Ti {\small II} &  3261.61 & 1.23 &  0.10 &  in Ti{\small II} 3261.58 &  in Ti{\small II} 3261.58  &    --  \\ 
  Ti {\small II} &   3271.65 &   1.24 &  -0.28 &    in Ti{\small II} 3272.07  &    --  &    --  \\ 
  Ti {\small II} &   3272.07 &   1.22 &  -0.19 &     90 (20)$^{b - Ti{\small II} + V{\small II}}$ &    --  &    --  \\ 
  Ti {\small II} &   3278.29 &   1.23 &  -0.32 &    30 (9) &    --  &    --  \\ 
  Ti {\small II} &   3278.92 &   1.08 &  -0.24 &    40 (12) &    --  &    --  \\ 
  Ti {\small II} &   3282.33 &   1.22 &  -0.33 &    36 (7) &    --  &    --  \\ 
  Ti {\small II} &   3287.65 &   1.89 &   0.46 &    75 (15) &   -- &    --  \\ 
  Ti {\small II} &   3318.02 &   0.12 &  -1.04 &    19 (4) &    --  &    --  \\ 
  Ti {\small II} &   3321.70 &   1.23 &  -0.31 &    42 (8) &    --  &    --  \\ 
  Ti {\small II} &   3322.93 &   0.15 &  -0.10 &   230 (50) &    72 (15)$^{b - Fe{\small II}}$  &    --  \\ 
  Ti {\small II} &   3326.76 &   0.11 &  -1.16 &    19 (4) &    --  &    --  \\ 
  Ti {\small II} &   3329.45 &   0.14 &  -0.26 &   148 (30) &   38 (10) &    --  \\ 
  Ti {\small II} &   3332.11 &   1.24 &  -0.11 &    42 (8) &    --  &    --  \\ 
  Ti {\small II} &   3335.19 &   0.12 &  -0.42 &   128 (26) &    48 (11)  &    --  \\ 
  Ti {\small II} &   3340.34 &   0.11 &	-0.54 &  120 (24)   &  23 (10)  &    -- \\  
  Ti {\small II} &   3341.87 &   0.57 &   0.35 &   330 (70) &    87 (14) &    18 (3) \\ 
  Ti {\small II} &   3349.03 &   0.61 &   0.43 &    in Ti{\small II} 3349.40 &   116 (14) &    13 (3) \\ 
  Ti {\small II} &   3349.40 &   0.05 &   0.53 &  1010 (200)$^{b - Ti{\small II}}$  &   180 (20) &    60 (5) \\ 
  Ti {\small II} &   3361.21 &   0.03 &   0.43 &   770 (150) &   148 (27) &    58 (5) \\ 
  Ti {\small II} &   3372.79 &   0.012 &   0.28 &   560 (110) &   109 (12) &    32 (4) \\ 
  Ti {\small II} &   3380.28 &   0.05 &  -0.63 &   120 (24) &    42 (13)  &    --  \\ 
  Ti {\small II} &   3383.76 &   0 &   0.16 &   450 (90) &   107 (21) &    31 (3) \\ 
  Ti {\small II} &   3387.83 &   0.03 & -0.41	& 127 (25) & 38 (13) & -- \\
  Ti {\small II} &   3394.57 &   0.012 &  -0.55 &   124 (25) &    27 (8)  &    --  \\ 
  Ti {\small II} &   3444.31 &   0.15 &  -0.65 &    54 (11) &    --  &    --  \\ 
  Ti {\small II} &   3461.50 &   0.14 &  -0.95 &    47 (9) &    --  &    --  \\ 
  Ti {\small II} &   3477.18 &   0.12 &  -0.96 &    32 (6) &    --  &    --  \\ 
  Ti {\small II} &   3491.05 &   0.11 & -1.15  &   35 (10)$^{b - Fe{\small I}}$  & -- & -- \\
  Ti {\small II} &   3504.89 &   1.89 &   0.39 &    80 (16) &    --  &    --  \\ 
  Ti {\small II} &   3510.84 &   1.89 &   0.29 &    50 (10) &   -- &    --  \\ 
  Ti {\small II} &   3535.41 &   2.06 &   0.03 &    30 (6) &    --  &    --  \\ 
  Ti {\small II} &  3685.19 &  0.57 &  -0.04 & in Ti{\small II} 3685.20 & in Ti{\small II} 3685.20 & in Ti{\small II} 3685.20 \\ 
  Ti {\small II} &   3685.20 & 0.61 &    0.13 &   380 (80)$^{b - Ti{\small II}}$ &   84 (6)$^{b - Ti{\small II}}$  &    20 (2)$^{b - Ti{\small II}}$ \\ 
  Ti {\small II} &   3741.64 &   1.58 &  -0.07 &    20 (4) &    --  &    --  \\ 
  Ti {\small II} &   3759.29 &   0.61 &   0.28 &   400 (100)$^{b - Fe{\small I}}$ &    72 (15) &    15 (2) \\ 
  Ti {\small II} &   3761.32 &   0.57 &   0.18 &   260 (50)$^{b - Fe{\small I}}$ &    67 (23) &    24 (3) \\ 
  Ti {\small II} &   3900.54 &   1.13 &  -0.20 &    31 (6) &    --  &    --  \\ 
  Ti {\small II} &   4468.51 &   1.13 &  -0.60 &    17 (5) &    --  &    --  \\ 
  Ti {\small II} &   4549.62 &   1.58 &  -0.11 &      36 (7) &   --  &    --  \\ 
   V {\small II} &   3125.28 &   0.32 &   0.04 &  in Cr{\small II} 3124.97 & no data &    --  \\ 
   V {\small II} &   3267.69	& 1.07 &	0.28	&	17 (6) & $<$ 80 & -- \\   
   V {\small II} &   3271.12  &   1.10 &   0.38 &     in Ti{\small II} 3272.07 &    --  &    --  \\   
   V {\small II} &   3276.12 &   1.13 &   0.49 &    11 (3) &    --  &    --  \\ 
  Cr {\small II} &   3124.97 &   2.46 &   0.30 &   180 (20)$^{b - V{\small II}}$ &  no data &    35 (15)  \\ 
  Cr {\small II} &	  3128.69 &   2.43 &	-0.54 &  30 (15) & no data & 10 (5)  \\
  Cr {\small II} &   3132.05 &   2.48 &   0.42 &   160 (30) &  no data  &    65 (10) \\ 
  Cr {\small II} &  3136.68	&  2.46  &  -0.45  &  30 (15) & no data & -- \\
  Cr {\small II} &   3147.22 &   2.48 &  -0.59 &    38 (10)  &    43 (15) &    --  \\ 
  Cr {\small II} &   3180.69  &  2.54 & -0.32 &  in Ca{\small II} 3179.33 &   in Ca{\small II} 3179.33 & -- \\
  Cr {\small II} &   3197.08 &   2.54 &  -0.43 &    42 (8) &    53 (20) &    --  \\ 
  Cr {\small II} &   3209.18 &   2.54 &  -0.56 &    --  &    57 (12) &    --  \\ 
  Cr {\small II} &   3217.39 &   2.54 &  -0.69 &    --  &    37 (11) &    --  \\ 
  Cr {\small II} &   3339.79 &   2.43 &  -0.89 &    --  &    20 (4) &    --  \\ 
  Cr {\small II} &   3342.58 &   2.46 &  -0.74 &    --  &    35 (14) &    --  \\ 
  Cr {\small II} &   3358.49 &   2.46 &  -0.59 &    10 (3) &    36 (13) &    --  \\ 
  Cr {\small II} &   3360.29 &   3.10 &  -0.32 &    --  &    32 (16) &    --  \\ 
  Cr {\small II} &   3368.04 &   2.48 &  -0.09 &    60 (12) &    82 (15) &    17 (3) \\ 
  Cr {\small II} &   3403.31 &   2.43 &  -0.67 &    30 (6) &    42 (6) &    --  \\ 
  Cr {\small II} &   3408.76 &   2.48 &  -0.39 &    33 (7) &    80 (9) &    22 (6) \\ 
  Cr {\small II} &   3421.20 &   2.42 &  -0.71 &    12 (4) &    50 (8) &    --  \\ 
  Cr {\small II} &   3422.73 &   2.46 &  -0.41 &    32 (6) &    72 (8) &    23 (7) \\ 
  Cr {\small II} &   3433.29 &   2.43 &  -0.73 &    25 (5) &    46 (15) &    --  \\ 
  Mn {\small II} &   3441.99 &   1.78 &  -0.36 &    21 (8) &    68 (9) &    20 (5) \\ 
  Mn {\small II} &   3460.31 &   1.81 &  -0.64 &    --  &    30 (9) &    --  \\ 
  Mn {\small II} &  3474.04 & 1.81 &  -0.93 & in Mn{\small II} 3474.13 &  in Mn{\small II} 3474.13 &   --  \\ 
  Mn {\small II} & 3474.13 & 1.83 & -1.06 & 10 (4)$^{b - Mn{\small II}}$ & 58 (11)$^{b - Mn{\small II}}$ &    --  \\ 
  Fe {\small I} &   3440.61 &   0 &  -0.67 &    39 (8)$^{b - Fe{\small I}}$ &    --  &    --  \\ 
  Fe {\small I} &   3440.99 &   0.05 &  -0.96 &  in Fe{\small I} 3440.61 &    --  &    --  \\ 
  Fe {\small I} &   3490.57 &   0.05 &  -1.10 &    in Ti {\small II} 3491.05  &    --  &    --  \\ 
  Fe {\small I} &   3565.38 &   0.96 &  -0.13 &    59 (12) &    --  &    --  \\ 
  Fe {\small I} &   3570.10 &   0.92 &   0.15 &   127 (25) &    45 (15)  &    --  \\ 
  Fe {\small I} &   3581.19 &   0.86 &   0.41 &   280 (60) &   107 (22) &    --  \\ 
  Fe {\small I} &   3608.86 &   1.01 &  -0.10 &    40 (8) &    --  &    --  \\ 
  Fe {\small I} &   3618.77 &   0.99 &  -0.00 &    87 (20)$^{b - Ni {\small I} }$ &    --  &    --  \\ 
  Fe {\small I} &   3631.46 &   0.96 &  -0.04 &    100 (20)$^{b - Sc{\small II}}$ &    47 (19) &    --  \\ 
  Fe {\small I} &   3719.93 &   0 &  -0.43 &   108 (22) &    46 (12) &    --  \\ 
  Fe {\small I} &   3734.86 &   0.86 &   0.32 &    in Ca{\small II} 3737 &    49 (15) &    --  \\ 
  Fe {\small I} &   3737.13	&  0.05  &	-0.57	 &   in Ca{\small II} 3737 &  --  & -- \\
  Fe {\small I} &   3745.56 &   0.09 &  -0.77 &    32 (6) &    --  &    --  \\ 
  Fe {\small I} &   3749.49 &   0.92 &   0.16 &   133 (27) &    55 (13) &    --  \\ 
  Fe {\small I} &   3758.23 &   0.96 &  -0.03 &   in Fe {\small I}  3759.29 & -- & -- \\
  Fe {\small I} &   3763.79 &   0.99 &  -0.24 &   in Fe {\small I}  3761.32 & -- & -- \\ 
  Fe {\small I} &   3815.84 &   1.49 &   0.24 &    47 (9) &    --  &    --  \\ 
  Fe {\small I} &   3820.43 &   0.86 &   0.12 &   130 (26) &    32 (10) &    --  \\ 
  Fe {\small I} &   3825.88 &   0.92 &  -0.04 &    52 (10) &    46 (15) &    --  \\ 
  Fe {\small I} &   3827.82 &   1.56 &   0.06 &  in 3827 interval &    --  &    --  \\ 
  Fe {\small I} &   3834.22 &   0.96 &  -0.30 & in 3827 interval &    --  &    --  \\ 
  Fe {\small I} &   3840.44 &   0.99 &  -0.51 & in 3827 interval &    --  &    --  \\ 
  Fe {\small I} &   3856.37 &   0.05 &  -1.29 &    32 (6) &    --  &    --  \\ 
  Fe {\small I} &   3859.91 &   0 &  -0.71 &    58 (12) &    --  &    --  \\ 
  Fe {\small I} &   4383.54 &  1.49  &  0.20  &  37 (8) & -- & -- \\
  Fe {\small II} &   3154.20   &   3.77 &  -0.51 &    48 (10)$^{b - Ti{\small II}}$ &   200 (40) &    25 (5) \\ 
  Fe {\small II} &   3167.86 &   3.81 &  -0.72 &   in Ti {\small II} 3168.5   &   100 (30) &    17 (7) \\ 
  Fe {\small II} &   3177.53 &   3.90 &  -0.90 &    --  &   100 (16) &    --  \\ 
  Fe {\small II} &   3183.11 &   1.70 &  -2.20 &    --  &    94 (46) &    --  \\ 
  Fe {\small II} &   3186.74 &   1.70 &  -1.77 &    57 (11) &    110 (30) &    10 (3)  \\ 
  Fe {\small II} &   3192.91 &   1.67 &  -2.01 &    30 (6) &    80 (12) &    14 (5) \\ 
  Fe {\small II} &   3193.80 &   1.72 &  -1.78 &    60 (12)$^{b - Fe{\small II}}$  &   120 (30)$^{b - Fe{\small II}}$  &    25 (10)$^{b - Fe{\small II}}$  \\ 
  Fe {\small II} &   3193.86 &  3.81 & -1.44 & in Fe{\small II} 3193.80 & in Fe{\small II} 3193.80 & in Fe{\small II} 3193.80 \\ 
  Fe {\small II} &   3196.07 &   1.67 &  -1.87 &    75 (15) &   103 (15) &    16 (3) \\ 
  Fe {\small II} &   3210.44 &   1.72 &  -1.76 &    43 (9) &   124 (22) &    33 (6) \\ 
  Fe {\small II} &   3213.31 &   1.70 &  -1.39 &   170 (30) &   200 (130) &    58 (6) \\ 
  Fe {\small II} &   3227.74 &   1.67 &  -1.18 &   300 (50) &   410 (60) &    90 (10) \\ 
  Fe {\small II} &   3237.82 &   3.89 &  -1.37 &  --   &   22 (6)  &  --  \\
  Fe {\small II} &   3243.72 &   4.15 &  -1.19 &    --  &    49 (17) &    --  \\ 
  Fe {\small II} &   3247.18 &   3.89 &  -1.17 &    15 (5) &    75 (53) &    --  \\ 
  Fe {\small II} &   3255.89 &   0.99 &  -2.56 &    20 (6) &    64 (17) &    --  \\ 
  Fe {\small II} &   3258.77 &   3.89 &  -1.15 &    10 (3) &    48 (13) &    --  \\ 
  Fe {\small II} &   3259.05 &   3.90 &  -0.97 &    16 (5) &    68 (24) &    --  \\ 
  Fe {\small II} &   3277.35 &   0.99 &  -2.39 &    20 (6) &    66 (14) &    --  \\ 
  Fe {\small II} &   3281.29 &   1.04 &  -2.74 &    13 (4) &    30 (8) &    --  \\ 
  Fe {\small II} &   3289.35 &   3.81 &  -1.57 &    --  &    24 (10) &    --  \\ 
  Fe {\small II} &   3323.06 &   3.97 &  -1.62 &    --  &    in Ti{\small II} 3322.93 &    --  \\ 
  Fe {\small II} &   4178.86 &   2.58 &  -2.51 &    --  &    21 (9) &    --  \\ 
  Fe {\small II} &   4233.17 &   2.58 &  -1.97 &    --  &    59 (18) &    --  \\ 
  Fe {\small II} &   4351.77 &   2.70 &  -2.25 &    --  &    31 (7) &    --  \\ 
  Fe {\small II} &   4522.63 &   2.84 &  -2.25 &    --  &    28 (10) &    --  \\
  Fe {\small II} &   4549.47 &   2.83 &  -2.09 &   --  &    55 (17) &    --  \\  
  Fe {\small II} &   4583.84 &   2.81 &  -1.93 &    --  &    50 (13) &    --  \\ 
  Fe {\small II} &   4923.93 &   2.89 &  -1.26 &    68 (14) &    77 (15) &    --  \\ 
  Fe {\small II} &   5018.44 &   2.89 &  -1.10 &    90 (18) &   142 (13) &    21 (7) \\ 
  Fe {\small II} &   5169.03 &   2.89 &  -1.00 &   113 (45) &   210 (20) &    42 (4) \\ 
  Fe {\small II} &   5234.62 &   3.22 &  -2.18 &    --  &    30 (5) &    --  \\ 
  Fe {\small II} &   5316.61 &   3.15 &  -1.87 &    --  &    70 (29) &    --  \\   
   Ni {\small I} &   3414.76 &   0.03 &  -0.014 &    30 (12) &    --  &    --  \\ 
  Ni {\small I} &   3524.53 &   0.03 &   0.008 &    40 (16) &    --  &    --  \\ 
  Ni {\small I} &   3619.39 &   0.42 &   0.04 &    in Fe {\small I} 3618.77 &    --  &    --  \\ 
    Ni {\small II} &   3513.99 &   2.87 &  -1.46 &    --  &    18 : &    --  \\    
   Sr {\small II} & 4077.71 &  0 	&	0.17   &    $<$ 60	& $<$ 90 & -- \\

interval & 3226-3231 & -- & -- & 450 (50)$^{b - Fe{\small II}+Ti{\small II}}$ &  see Fe{\small II}, Ti{\small II} & see Fe{\small II}  \\
interval &	3800--3850 & -- & --  &	2,100 (400)$^{b - Mg{\small I}+Fe{\small I}}$  & see Mg{\small I}, Fe{\small I} & see Mg{\small I}  \\

\label{tab:linelist}
\end{longtable}
\end{center}
$^a$ Laboratory wavelength in air. Wavelength, $\chi$ (excitation energy of lower state) and log $gf$ data from the Vienna Atomic Line Database. \\ 
$^b$ Blended line - contributing element noted in italic with line data given in the table (Table \ref{tab:linelist}). The equivalent width is from the total summed flux of the blend. \\
$^c$ G241-6 abundances from \citet{zuckerman2010} are associated with the EWs of this table.  Spectra of G241-6 are shown in Figure \ref{fig:g241-6_FeMgSi} of this paper and Figure 4 of \citet{zuckerman2010}. \\
$^d$ Combined Ca H\&K EW measurement from \citet{koester2005} is used since the HIRES data do not fully cover the red wing of this large feature.  \\

\begin{table}[htdp]
\caption{White Dwarf Models, Derived Masses and Accretion Rates}
\begin{center}
\begin{tabular}{l c c c c c c l}
\hline
\hline
	&   T$_{eff}$ & log $g$ &  M$_*$ & log M$_{cvz}^{He}$/M$_*$ & M$_{cvz}^{high-Z}$ & $<\dot{M}_{acc}>$ \\
	&	(K)	  &  (cgs)  &  ($\Msun$)	&			& (10$^{22}$g)  & (g s$^{-1}$)  \\
\hline
PG1225 model 1 & 10500 & 7.7 & 0.42 &     -4.30     &  5.6 & 3 $\times$ 10$^8$   \\
PG1225 model 2 & 10800 & 8.0 & 0.58 &     -5.07     &  2.1 &  5 $\times$ 10$^8$  \\
PG1225 model 3 & 11100 & 8.3 & 0.77 &     -5.84     &  0.7 & 9 $\times$ 10$^8$  \\
\hline
HS2253 model 1 & 14000 & 8.1 & 0.65 &     -5.77     &  6.0 & 5 $\times$ 10$^9$   \\
HS2253 model 2 & 14400 & 8.4 & 0.84 &     -6.48     &  2.0 &  9 $\times$ 10$^9$  \\
HS2253 model 3 & 14800 & 8.7 & 1.03 &     -7.21     &  0.6 &  1.4 $\times$ 10$^{10}$    \\
\hline

\end{tabular}
\end{center}
\label{tab:modelsmass}
Convective zone (CVZ) helium masses, contaminant (high-Z) masses, and time-averaged mass accretion rates (excluding hydrogen) for WD models of varying effective temperature (T$_{eff}$) and gravity ($g$).  The (steady state) accretion rates are the sum of individual element rates calculated from the mass of the element in the CVZ divided by its settling time, with abundances and settling times from Tables \ref{tab:hs2253_abund} and \ref{tab:pg1225_abund}.  For PG1225-079 the Si upper limit is used, and an abundance of oxygen is assumed as would derive from the dominant mineral oxides: MgO, SiO$_2$, FeO, CaO.  Including carbon at its upper limit \citep{wolff2002} increases the high-Z masses and accretion rates by $\simeq$ 30\% for PG1225-079, and is negligible for HS2253+8023.
Including aluminum at its upper limit increases the high-Z masses and accretion rates by $\simeq $5\% for both stars.
\end{table}%

\begin{table}[htdp]
\caption{HS2253+8023 Atmospheric Abundances by Number and Settling Times for 3 Different T$_{eff}$/log $g$ Models}
\begin{center} 
\begin{tabular}{l c c c c c c c l}
\hline
\hline
  & 14000/8.1 & 14400/8.4 & 14800/8.7 & 14000/8.1 & 14400/8.4 & 14800/8.7  \\ 
Z	&	log(Z/He)  & log(Z/He) & log(Z/He)  &  log($\tau$/yrs) & log($\tau$/yrs)  & log($\tau$/yrs)   \\
\hline
H & -5.70 (0.06) & -5.62 (0.06) & -5.55 (0.06)  &	--	&	--	&	--  \\ 
O & -5.48 (0.07) & -5.37 (0.07) & -5.28 (0.07)	&	5.71	&	5.01	&	4.31 \\ 
Mg & -6.23 (0.06) & -6.10 (0.04) & -6.02 (0.03)	&	5.63	&	4.93	&	4.23 \\ 
\textsl{ from Mg {\small I} } & $\;$  \textsl{ -6.27 (0.05) } & $\;$ \textsl{ -6.12 (0.06) } &  $\;$ \textsl{ -6.03 (0.04) } & -- & -- & -- \\
\textsl{ from Mg {\small II} } & $\;$ \textsl{ -6.17 (0.05) } & $\;$ \textsl{ -6.07 (0.05) } &$\;$ \textsl{ -6.01 (0.05) } & -- & -- & -- \\
Si & -6.40 (0.03) & -6.27 (0.03) & -6.16 (0.04)	&	5.63	&	4.93	&	4.21	 \\ 
Ca & -7.10 (0.03) & -6.99 (0.03) & -6.92 (0.05)	&	5.50	&	4.78	&	4.06 \\ 
Ti & -8.92 (0.02) & -8.74 (0.02) & -8.55 (0.02)	&	5.43	&	4.73	&	4.03 \\ 
Cr & -8.17 (0.04) & -8.01 (0.03) & -7.83 (0.03)	&	5.44	&	4.74	&	4.01 \\ 
Mn & -8.63 (0.09) & -8.42 (0.10) & -8.24 (0.09)	&	5.43	&	4.71	&	3.99 \\ 
Fe & -6.34 (0.04) & -6.17 (0.03) & -6.00 (0.02)	&	5.42	&	4.71	&	3.98 \\ 
\textsl{ from {Fe \small I} } &$\;$ \textsl{ -6.29 (0.05) } &$\;$ \textsl{  -6.15 (0.04) } &$\;$ \textsl{ -5.99 (0.05)  } & -- & -- & -- \\
\textsl{ from {Fe \small II} } &$\;$ \textsl{ -6.36 (0.03) } &$\;$ \textsl{ -6.18 (0.03) } &$\;$ \textsl{  -6.01 (0.03)  } & -- & -- & -- \\
Ni & -7.51 : & -7.31 : & -7.14 : 				&	5.40	&	4.68	&	3.96	\\
C &	--	&		--		&	--		&	5.74	&	5.05	&	4.35 \\
Na	&	--	& $<$  -6.8 & -- &    	5.63	&	4.92	&	4.21	\\
Al	& 	--	& $<$ -6.7 &  -- 	&	5.61	&	4.92	&	4.22	\\
Sc	&	--	& $<$ -9.8 & -- & 	5.45	&	4.74	&	4.03	\\
V	&	--	& $<$ -8.6 & -- &  	5.42	&	4.73	&	4.02	\\
Sr	&	--	& $<$ -9.6 & -- &	5.24	&	4.53	&	3.81 \\

\hline
\\  
\end{tabular}
\end{center}
\label{tab:hs2253_abund}

Element abundances and settling times derived from the equivalent width measurements of Table \ref{tab:linelist}  and different models, as described in Sections \ref{sec:models}, \ref{sec:abundances}, and \ref{sec:accretiondiffusion}.
Uncertainties in parentheses are either the standard deviation of the mean for the set of individual line abundances, or the propagated measurement error, whichever is greater.  When a significant difference is measured between ionization states of an element (only true for iron), this additional uncertainty is added in.  While the absolute abundances and diffusion times vary considerably between models, the relative abundances and relative diffusion times among the listed elements are far less sensitive to changes in the model.  
  \\

\end{table}%

\begin{table}[htdp]
\caption{PG1225-079 Atmospheric Abundances by Number and Settling Times for 3 Different T$_{eff}$/log $g$ Models}
\begin{center} 
\begin{tabular}{l c c c c c c l}
\hline
\hline

	&	10500/7.7  & 10800/8.0  & 11100/8.3  &  10500/7.7  & 10800/8.0  & 11100/8.3\\
Z	&	log(Z/He)  & log(Z/He) & log(Z/He)  &  log($\tau$/yrs) & log($\tau$/yrs)  & log($\tau$/yrs)   \\
\hline

H &  -4.13 (0.10) & -4.05 (0.10) & -3.99 (0.10) &	--	&	--	&	-- \\
Mg &  -7.44 (0.05) & -7.27 (0.05) & -7.11 (0.05) & 6.85 & 6.14 & 5.41 \\
\textsl{ from Mg {\small I} } & \textsl{ -7.44 (0.06)} & \textsl{ -7.27 (0.06)} & \textsl{ -7.11 (0.06)} & -- & -- & -- \\
\textsl{ from Mg {\small II} } & \textsl{ -7.43 (0.10)} & \textsl{ -7.30 (0.10)} & \textsl{ -7.11 (0.10)} & -- & -- & -- \\
Ca &  -8.25 (0.03) & -8.06 (0.03) & -7.90 (0.04) & 6.83	& 6.07	& 5.32 \\
\textsl{ from Ca {\small I} }  &  \textsl{ -8.30 (0.15)} & \textsl{ -8.18 (0.15)} & \textsl{ -8.04 (0.15)} & -- & -- & -- \\
\textsl{ from Ca {\small II} } &  \textsl{ -8.24 (0.03)} & \textsl{ -8.04 (0.03)} & \textsl{ -7.89 (0.04)} & -- & -- & -- \\
Sc &  -11.57 (0.07) & -11.29 (0.07) & -11.05 (0.07) & 6.78 & 6.03	&	5.28 \\
Ti &  -9.68 (0.02) & -9.45 (0.02) & -9.23 (0.02) & 6.76	&	6.01	& 5.26\\
V	& -10.68 (0.10)	& -10.41 (0.10)	& -10.18 (0.10) & 6.73 & 5.98 &	5.23	\\
Cr &  -9.47 (0.06) & -9.27 (0.06) & -9.05 (0.06) & 6.72 &	5.97 &	5.24\\
Mn &  -10.03 (0.14) & -9.79 (0.14) & -9.52 (0.14) & 6.70 &	 5.96	 &  5.24 \\ 
Fe &  -7.64 (0.03) & -7.42 (0.07) & -7.21 (0.08) &  6.69 &	5.97 &	5.25 \\
\textsl{ from {Fe \small I} }  &  \textsl{  -7.67 (0.03)} & \textsl{  -7.50 (0.03)} & \textsl{  -7.30 (0.03)} & -- & -- & -- \\
\textsl{ from {Fe \small II} } &  \textsl{  -7.62 (0.03)} & \textsl{  -7.36 (0.03)} & \textsl{  -7.15 (0.03)} & -- & -- & -- \\
Ni &  -8.88 (0.14) & -8.76 (0.14) & -8.64 (0.14) & 6.70 & 5.99 & 5.24  \\
C  &	--		&	--		&	--	  & 6.89  &  6.18  &  5.47 \\
O & $<$ -5.69    & $<$ -5.54    & $<$ -5.44  & 6.86 & 6.15	& 5.43   \\
Na & $<$ -8.45    & $<$ -8.26    & $<$ -8.02  & 6.83  &  6.12  &  5.41 \\
Al & $<$ -8.04    & $<$ -7.84    & $<$ -7.68  &   6.84  &	6.11  &  5.38 \\
Si & $<$ -7.49    & $<$ -7.27    & $<$ -7.07 &   6.85 & 6.11 & 5.39 \\
Sr & $<$ -11.90    & $<$ -11.65    & $<$ -11.38  &  6.56	& 5.85 &	5.10 \\
\hline
\\
\end{tabular}
\end{center}
\label{tab:pg1225_abund}

Similar to Table \ref{tab:hs2253_abund}, but for PG1225-079. \\

\end{table}%

\begin{table}[htdp]
\caption{Abundance Ratios By Number}
\begin{center}
\begin{tabular}{l c c c c}
\hline
\hline
		&	PG 1225-079			&	HS 2253+8023		&	Solar	\\
		&	Convective Zone	&	Convective Zone	& \citep{lodders2003}	\\
\hline
H/Mg		&	1700 (500) 		&	3.1	(0.6)			&	$2.8 \times10^4$	\\
C/Mg	&	$<$ 5 $^a$			&	$<$	0.01	$^b$		&	6.9	\\
O/Mg		&	$<$ 60			&		5.4	(1.1)   	&	14	\\
Na/Mg		&	$<$ 0.12			&		$<$ 0.22		&	0.056  \\
Al/Mg		&	$<$ 0.31			&		$<$	0.30		&	0.081	\\
Si/Mg		&	$<$ 1.1			&		0.68	(0.10)	&	0.98 	\\
Ca/Mg		&	0.16 (0.02) 		&		0.13	(0.02)	&	0.062	\\
Sc/Mg		&	$9.4 \pm  2.7 \times10^{-5}$	&  $<  2.2 \times10^{-4}$	& $3.3 \times10^{-5}$  \\  
Ti/Mg	& $6.7 \pm 1.2 \times10^{-3}$  &  $2.4 \pm 0.6  \times10^{-3}$ & $2.3  \times10^{-3}$ \\
V/Mg		&	$7.1 \pm 2.2 \times10^{-4}$   &	 $< 3.7 \times10^{-3}$ &	$2.8  \times10^{-4}$ \\ 

Cr/Mg	&	0.010 (0.002) 		&		0.013	(0.003) 	&	0.013	\\
Mn/Mg	&  $3.1 \pm 1.2 \times10^{-3}$  & $5.1 \pm 1.7 \times10^{-3}$ & $8.9 \times10^{-3}$ \\
Fe/Mg		&	0.71 (0.12) 	&		0.89	(0.18)	&	0.83	\\
Ni/Mg		&	0.033 (0.011)		&		0.06 :  & 0.047	 \\
Sr/Mg		&	$<$ 4.5$\times10^{-5}$  &	 $<$ 3.1$\times10^{-4}$  &	$2.3  \times10^{-5}$	\\ 
\hline

\end{tabular}
\end{center}
\label{tab:abund_ratios}
$^a$ C/Mg from Wolff et al. (2002). \\
$^b$ C/Mg from Friedrich et al. (1999). \\

Measured atmospheric abundance ratios calculated as the average of 3 models with ratios from Tables \ref{tab:hs2253_abund} and \ref{tab:pg1225_abund}.    Uncertainties in parentheses are a combination of the propagated abundance ratio measurement uncertainties, and the root-mean-square of ratio values from different models from Tables \ref{tab:hs2253_abund} and \ref{tab:pg1225_abund}, added in quadrature.  \end{table}%

\clearpage

\begin{figure}[htbp]
\begin{center}
  \includegraphics[width=140mm]{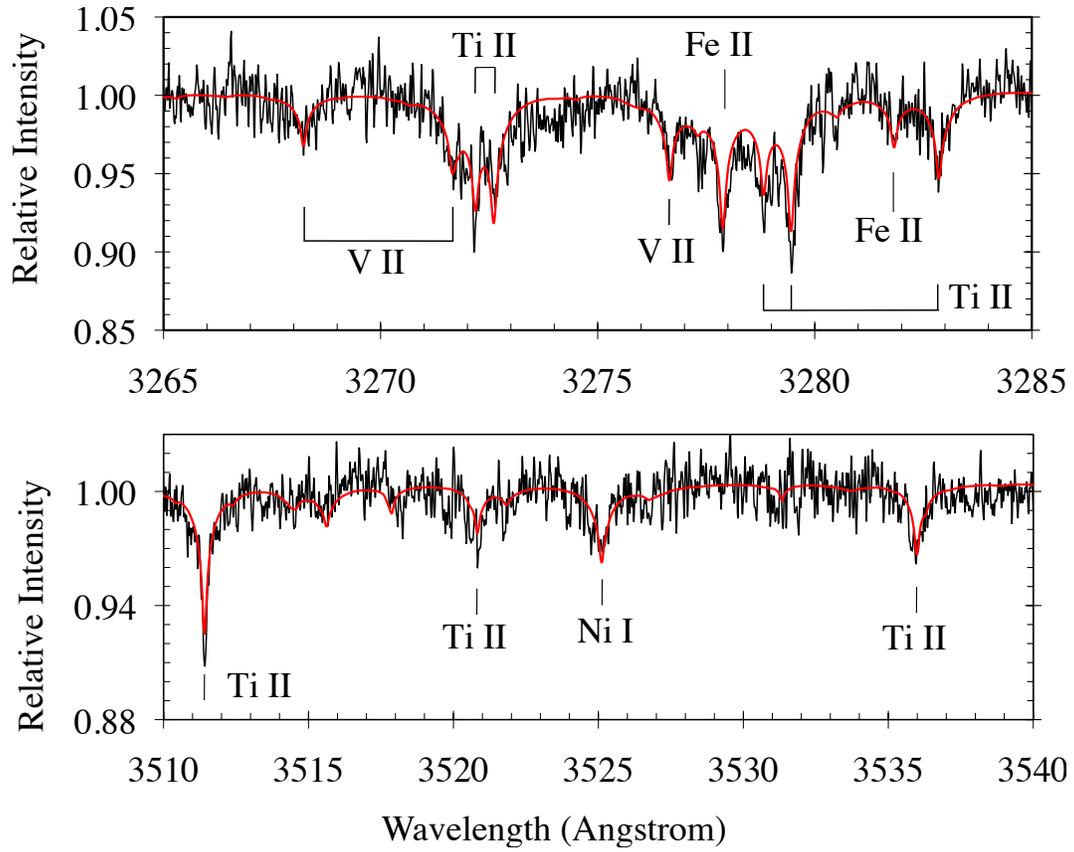}
\caption{PG1225-079. Similar to Figure \ref{fig:pg1225_MgI}, but for V and Ni.  The data in the lower panel are smoothed by a 3 point boxcar average.  }
\label{default}
\end{center}
\end{figure}

\begin{figure}[htbp]
\begin{center}
  \includegraphics[width=140mm]{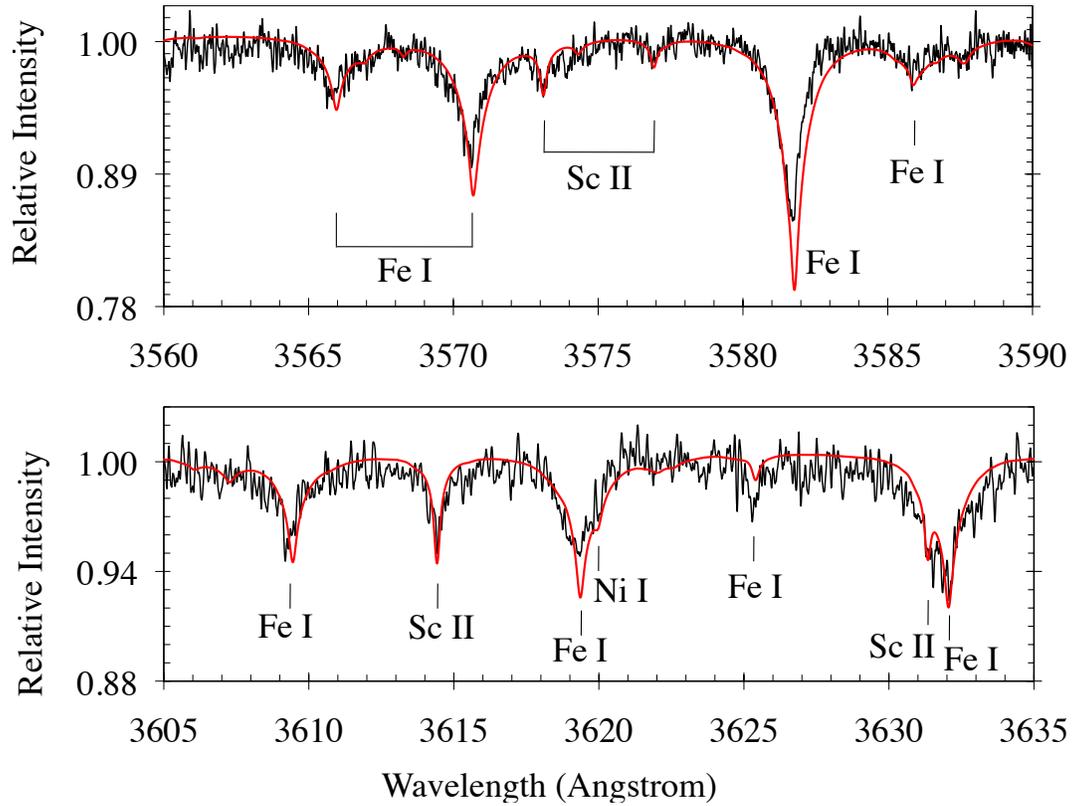}
\caption{PG1225-079. Similar to Figure \ref{fig:pg1225_MgI}, but smoothed by a 3-point (upper panel) and 5-point (lower panel) boxcar average, and displaying Sc {\small II} and Fe {\small I} lines; a probable blend of a Ni {\small I} feature is also labeled.  The apparent ``mis-fit'' of some Fe {\small I} lines is due to a combination of the $\sim$ 0.1 dex discrepancy of Fe {\small I} and Fe {\small II} abundances, uncertainties in the modeled line data, and an un-modeled shift of Fe {\small I} features (see also Figure \ref{fig:vel_hist}).}
\label{fig:pg1225_FeI_Sc}
\end{center}
\end{figure}

\begin{figure}[htbp]
\begin{center}
  \includegraphics[width=140mm]{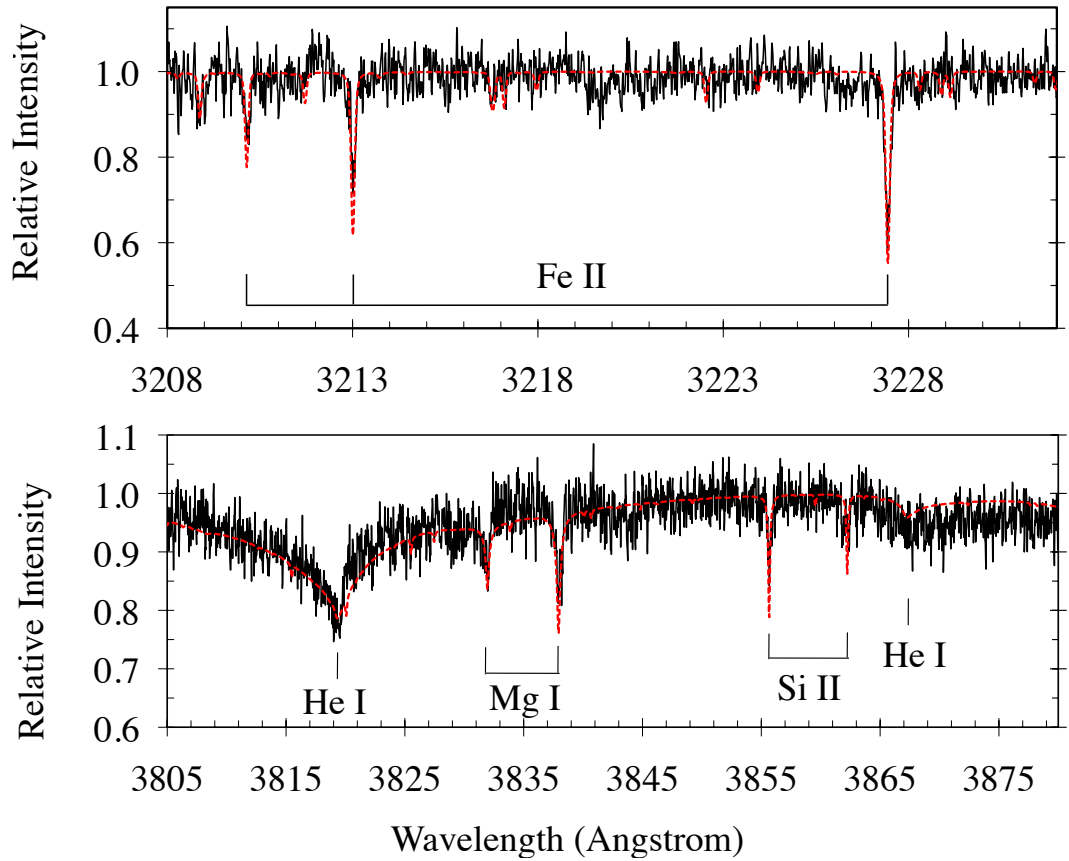}
\caption{Unsmoothed portions of the combined HIRES spectrum of G241-6 is plotted in black, displaying Fe {\small II}, Mg {\small I}, Si {\small II} and He {\small I} lines.  Wavelengths are in air and a heliocentric rest frame.  Plotted in red is a T$_{eff}$=15,300 K log $g$ = 8.0 model,  convolved with a Gaussian to approximate the instrumental resolution, and with element abundances from \citet{zuckerman2010}; the model is blue shifted by -28.0 km s$^{-1}$, the mean velocity from the full set of measured absorption lines.  See also Figure 4 of \citet{zuckerman2010} for a display of O {\small I} lines in G241-6.}
\label{fig:g241-6_FeMgSi}
\end{center}
\end{figure}

\begin{figure}[htbp]
\begin{center}
  \includegraphics[width=120mm]{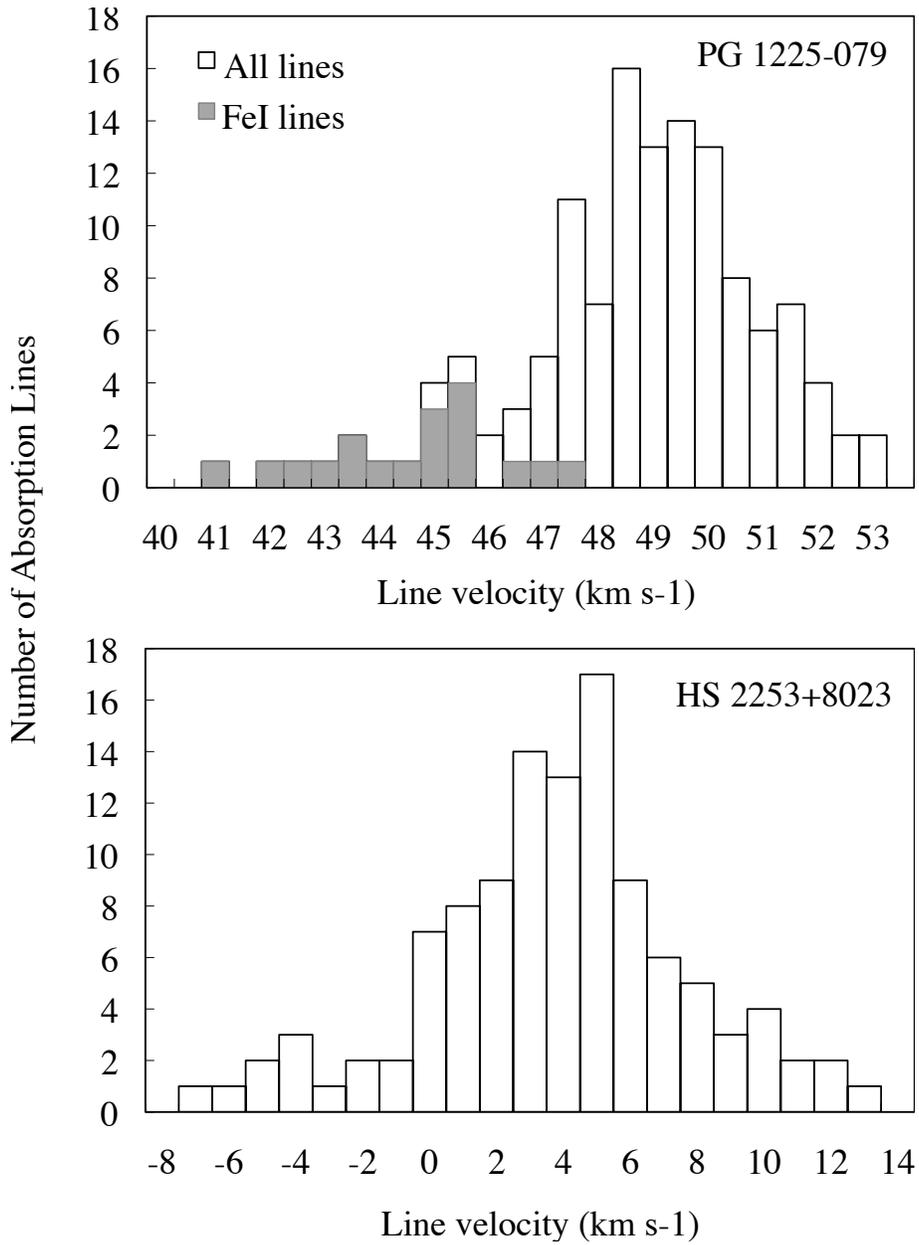}
\caption{Histograms of (heliocentric) velocity measurements from HIRES data.  The velocity bin size is 0.5 km s$^{-1}$ for PG1225 and 1 km s$^{-1}$ for HS2253;  the larger velocity dispersion for the latter star is likely due to the lower SNR.   Fe {\small I} lines are a distinguished set in PG1225, whereas the eight Fe {\small I} lines in HS2253+8023 span between -3 km s$^{-1}$ and +12 km s$^{-1}$.}
\label{fig:vel_hist}
\end{center}
\end{figure}


\begin{figure}[htbp]
\begin{center}
  \includegraphics[width=120mm]{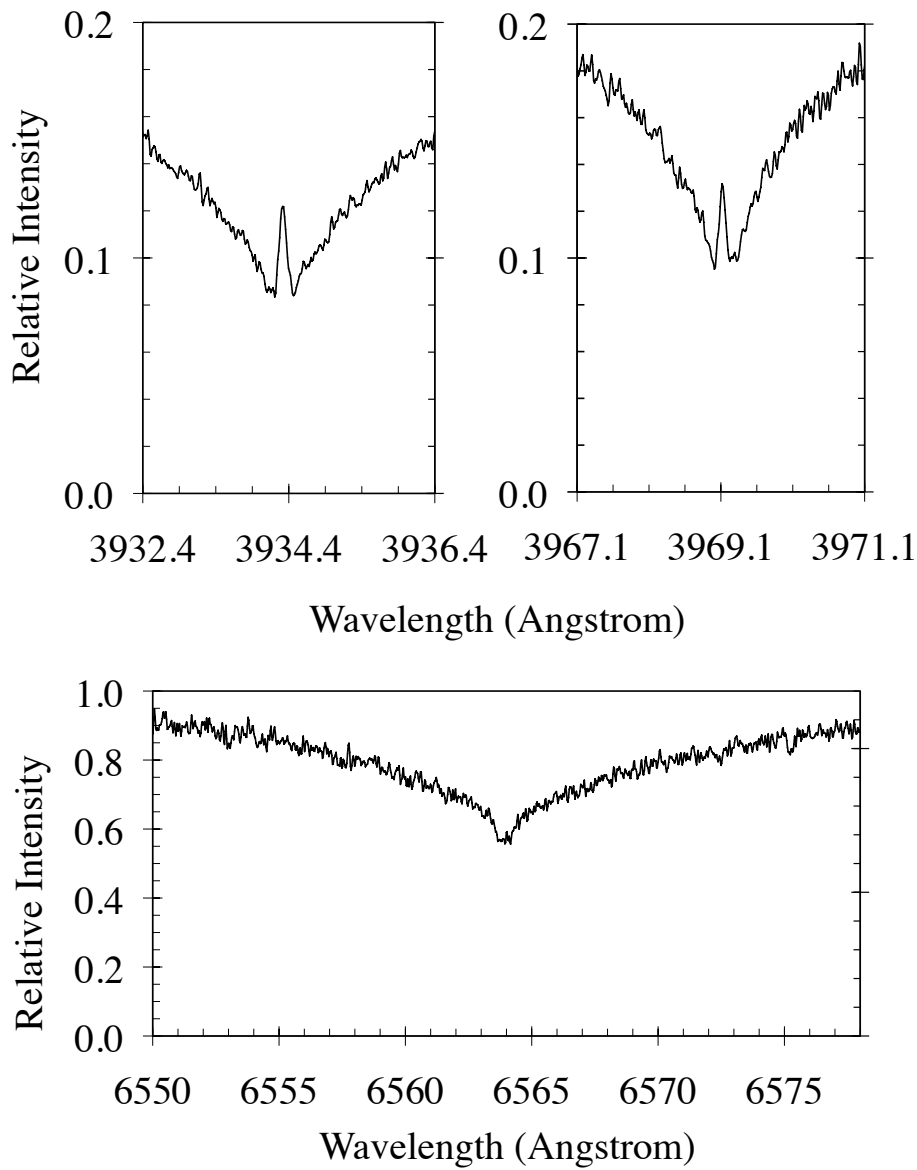}
\caption{Upper Panel: HIRES observations of Ca {\small II} H \& K emission cores in the white dwarf PG1225-079; zoomed-in display of Figure \ref{fig:pg1225_blu}.  The local level of the stellar continuum is near 0.55 on this ordinate axis of arbitrary scale.  The equivalent widths of the core emission features, relative to the lowest depth of the associated calcium lines, are $\sim$50 m\AA\ centered on the peak at $\lambda$3934.32 \AA\ and $\sim$40 m\AA\ centered on the peak at $\lambda$3969.12 \AA.  Lower Panel:  Core portion of the H$\alpha$ profile of PG1225-079.  All wavelengths are in air and a heliocentric rest frame. }
\label{fig:pg1225_CaHK}
\end{center}
\end{figure}

\begin{figure}[htbp]
\begin{center}
  \includegraphics[width=120mm]{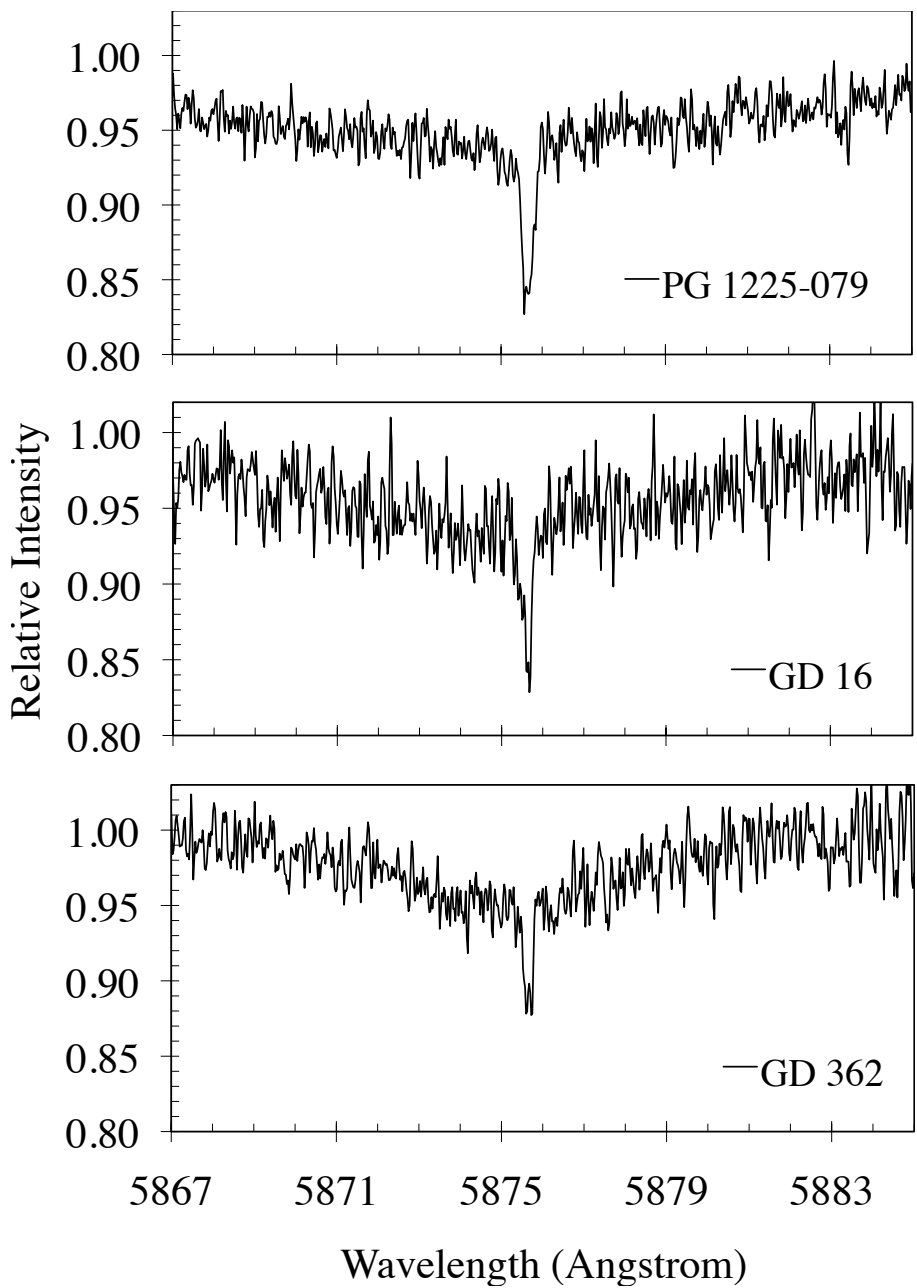}
\caption{Narrow cores on the broad He {\small I} $\lambda$5876 line are detected in HIRES spectra of three helium-dominated, hydrogen-lined, externally polluted WDs in the T$_{eff}$ range of 10,000 K to 12,000 K.  PG1225-079's observations are from Table \ref{tab:obs}, GD16 was observed in both February and November of 2008 with no noted variations of the He line, and GD362's observations are from \citet{zuckerman2007}.  The equivalent widths of the core features are:  50 m\AA, 45 m\AA, and 25 m\AA\ in PG1225-079, GD16, and GD362, respectively.  For this display each WD spectrum was Doppler shifted to the laboratory rest frame, by its measured redshift, to show the alignment of the core feature with the central wavelength in air $\lambda$5875.70 \AA\ of the helium line.}

\label{fig:3He5876cores}
\end{center}
\end{figure}

\begin{figure}[htbp]
\begin{center}
  \includegraphics[width=120mm]{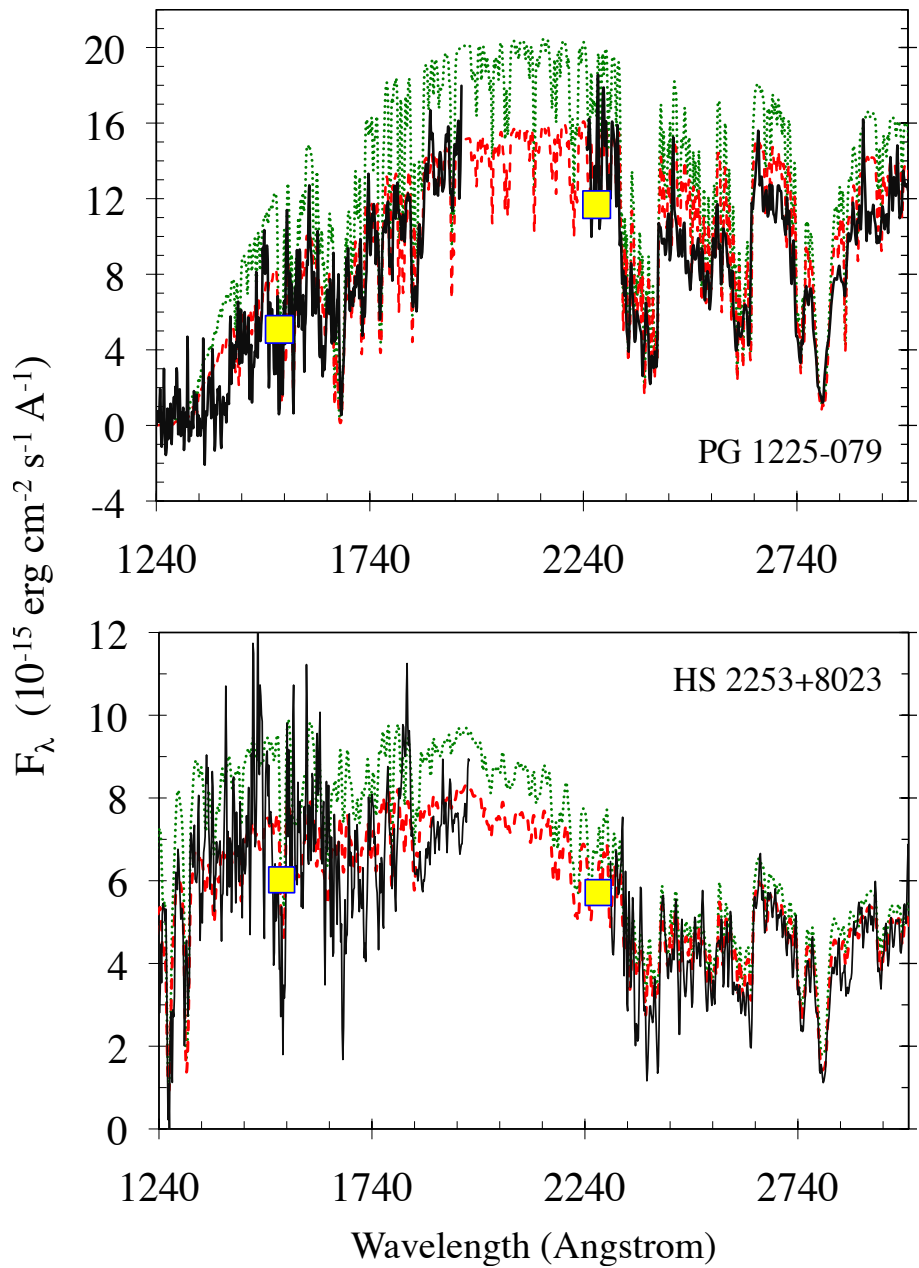}
\caption{Portions of the spectral energy distributions (SED) of PG1225-079 (top) and HS2253+8023 (bottom) displaying archival IUE spectra (black, solid line) and GALEX photometry (yellow boxes). The models are convolved with 6\AA\ Gaussians and scaled to match visual photometry.  For PG1225-079 the 10,500 K model is a red, dashed line, and the 11,100 K model is a green, dotted line.  For HS2253+8023 the 14,000 K model is a red, dashed line, and the 14,800 K model is a green, dotted line.    An extensive list of UV absorption lines is included in the models, with element abundances from the current HIRES measurements (Tables \ref{tab:hs2253_abund} and \ref{tab:pg1225_abund}) and carbon from previous UV analyses \citep{friedrich1999,wolff2002}.   In both cases the lower temperature model fits the SED well, but if there is some unaccounted-for extinction, then the higher temperature model may be appropriate. }
\label{fig:IUE_fit}
\end{center}
\end{figure}

\begin{figure}[htbp]
\begin{center}
  \includegraphics[width=140mm]{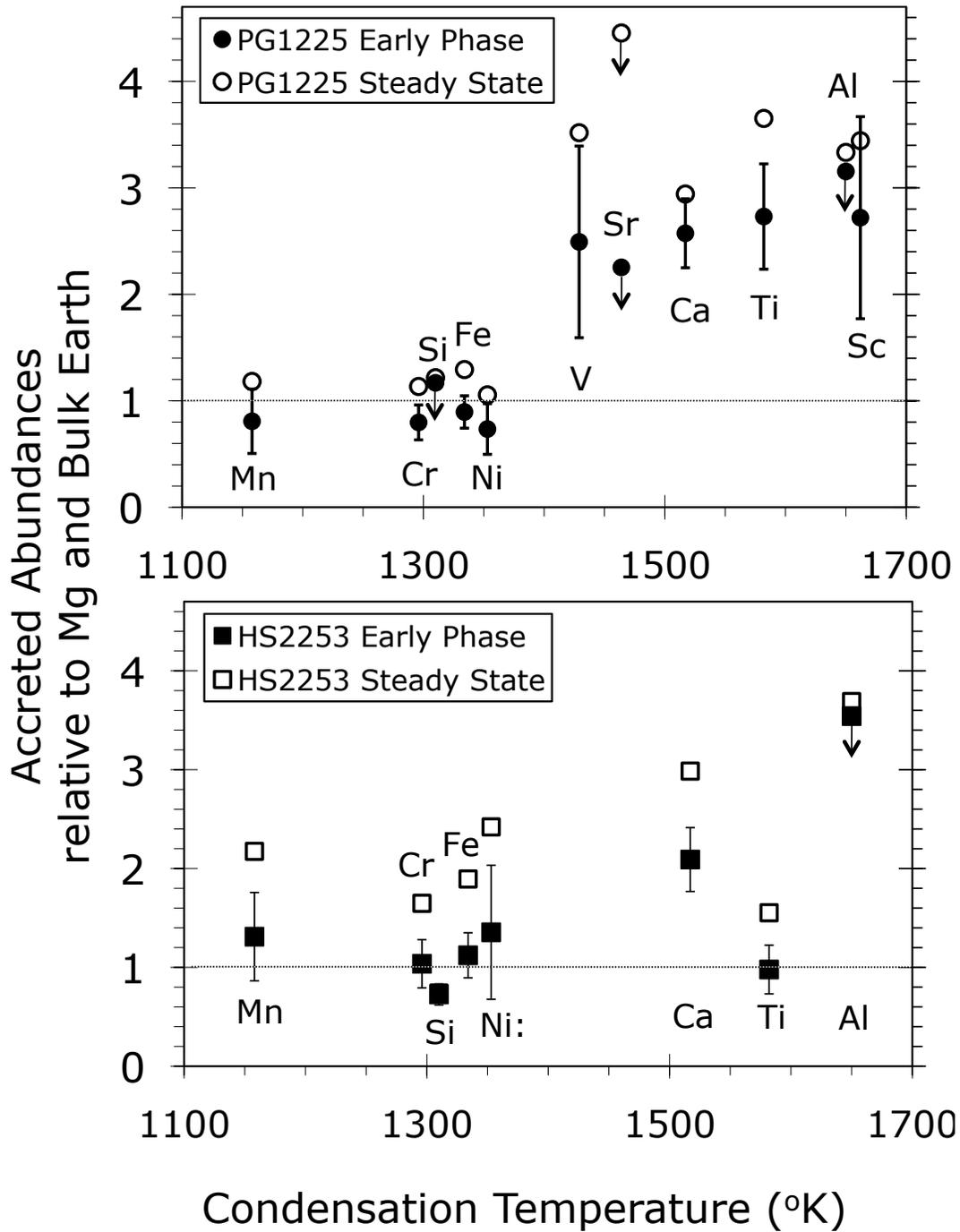}
\caption{Parent body abundance ratios from Table \ref{tab:abund_ratios}, with values for Earth from \citet{allegre2001} and settling times from Tables \ref{tab:hs2253_abund} and \ref{tab:pg1225_abund}; 50\% condensation temperatures are from \citet{lodders2003}.  The fiducial element, Mg, has a 50\% T$_c$ = 1336 K, nearly the same as that of Fe.  Oxygen is off the plot at 50\% T$_c$ = 180 K. The dotted line at ordinate value 1 is to guide the eye to Earth-like values.}
\label{fig:cond_temp}
\end{center}
\end{figure}

\newpage

\begin{figure}[htbp]
\begin{center}
  \includegraphics[width=140mm]{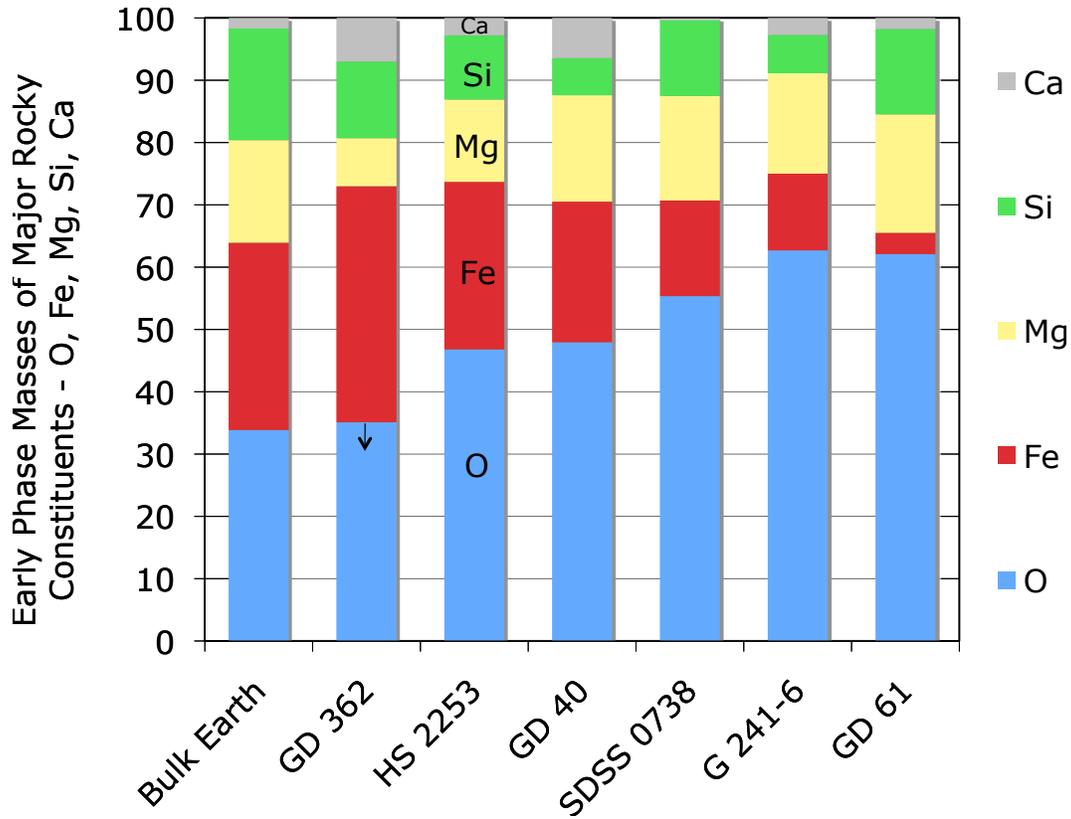}
\caption{Accreted parent body mass compositions of major elements not including hydrogen, for an early phase interpretation.  For bulk Earth, HS 2253, GD 40, G 241-6 and SDSS 0738, the omission of hydrogen has a negligible effect on this display, since it is less than 1\% of the bulk mass in each of these systems.  Conversely, GD 362 and GD 61 have large amounts of hydrogen in their atmospheres, which may or may not be associated with currently observed high-Z elements from polluting parent bodies;  see Section \ref{sec:the_set}.  Abundance data for the bulk Earth are from \citet{allegre2001}; GD362 from \citet{zuckerman2007}; GD40 from \citet{klein2010}; SDSS0738 from \citet{dufour2010}; G241-6 from \citet{zuckerman2010}; GD61 from \citet{farihi2011gd61}. }
\label{fig:bar_graph_early}
\end{center}
\end{figure}

\begin{figure}[htbp]
\begin{center}
  \includegraphics[width=140mm]{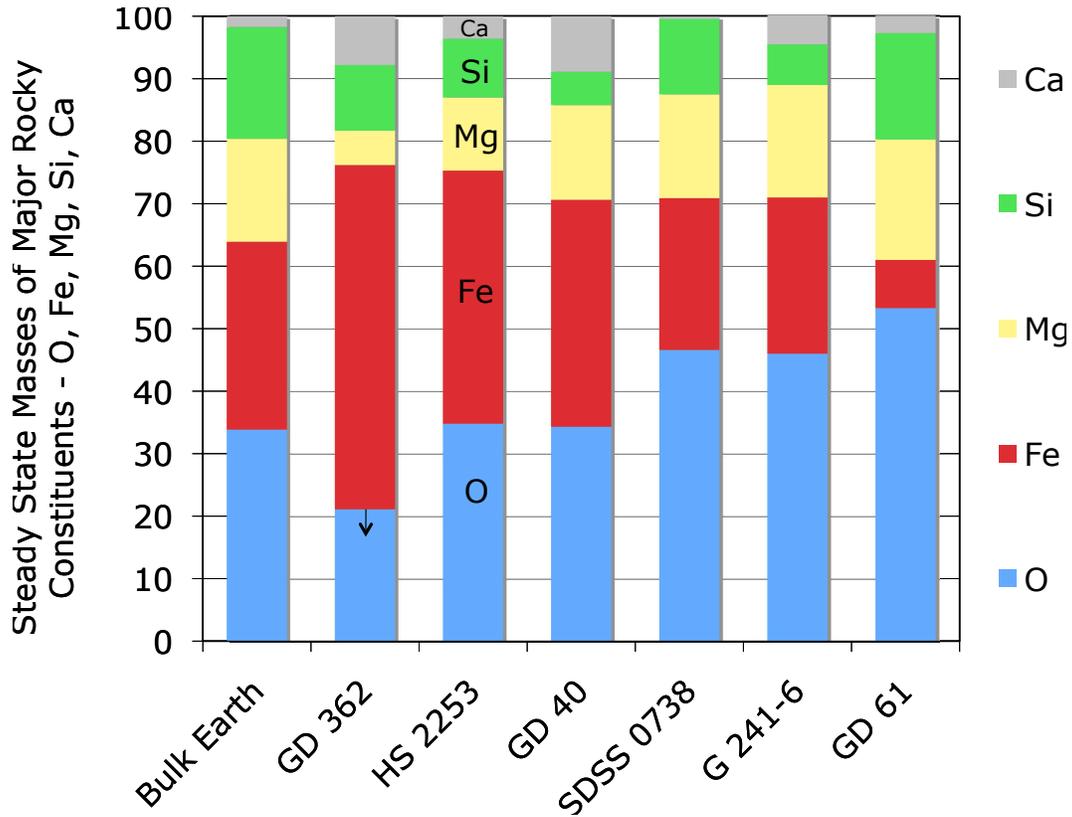}
\caption{Similar to Figure \ref{fig:bar_graph_early}, but here in the steady state interpretation of parent body composition.  Abundances for white dwarf systems are from the same sources as in Figure \ref{fig:bar_graph_early}, with the element abundance ratios scaled by ratios of settling times.  Settling time data for GD362 are from \citet{koester2009}; HS2253 from Table \ref{tab:hs2253_abund}; GD40 from \citet{klein2010}; SDSS0738 from our calculations analogous to \citet{koester2009}; G241-6 are the same as GD40; and GD61 from \citet{farihi2011gd61}.  } 
\label{fig:bar_graph_steady}
\end{center}
\end{figure}

\end{document}